\documentstyle[psfig]{l-aa}

\newcommand{\mincir}{\raise -2.truept\hbox{\rlap{\hbox{$\sim$}}\raise5.truept
\hbox{$<$}\ }}
\newcommand{\gr}{\kern 2pt\hbox{}^\circ{\kern -2pt K}} 
\newcommand{\magcir}{\raise -2.truept\hbox{\rlap{\hbox{$\sim$}}\raise5.truept
\hbox{$>$}\ }}
\newcommand{\oml}{\Omega_{\Lambda}}
\newcommand{\apj}{ApJ}
\newcommand{\apjl}{ApJL}
\newcommand{\mnras}{MNRAS}
\newcommand{\aap}{AA}
\newcommand{\apjs}{ApJS}

\begin{document}

\thesaurus{
       (12.03.4;  
        12.04.1;  
        12.12.1;  
Universe
        11.03.1;  
          }

\title{ Large scale structure formation  in mixed dark matter models
  with a cosmological constant}

\author{R.~Valdarnini \inst{1}, T.~Kahniashvili \inst{2},
B.~Novosyadlyj \inst{3} }

\offprints{Riccardo Valdarnini}

\institute{SISSA, via Beirut 2-4, 34014, Trieste, Italy
\and Abastumani Astrophysical Observatory, A.Kazbegi ave.2a, 380060,
Tbilisi, Georgia
\and Astronomical Observatory of L'viv State University, Kyryla and
Mephodia str.8, 290005, L'viv, Ukraine}

\date{Received \dots; accepted \dots}
\maketitle
\markboth{ Mixed Dark Matter Models
 with a Cosmological Constant}{}

\begin{abstract}
We study linear power spectra and formation of large scale structures
 in flat cosmological models with
$\Lambda \ge 0$ and cold plus hot dark matter components.
We refer to these models as mixed $\Lambda$ models (MLM).
The hot component consists of massive neutrinos with
 cosmological density $\Omega_H$ 
 and number of neutrino species as a free parameter.
The linearized Einstein-Boltzmann
equations for the evolution of the metric and density perturbations are
integrated for a set of values of the cosmological parameters.
We  study MLM models with present matter density in the range 
$ 0.25 \le \Omega_M \le 1$, dimensionless Hubble
constant $0.4 \le h\le 0.7$ and the hot dark matter content with 
 a ratio within the limits $ 0 \le \Omega_H/\Omega_M\le0.3$.
For all the considered models we assume a scale-invariant primeval spectrum.

The density weighted final linear power
spectra are normalized to the four year COBE data and have been used to 
constrain
the parameter space  by a comparison of  linear predictions with the 
current observational data on large scales.
 The consistency
of MLM predictions with the observable data set is best obtained for models
with
one species of massive neutrinos and $\Omega_H/\Omega_M\le0.2$. 
 Of the considered linear tests the strongest constraints on $\Omega_M$
 that we obtain arise by comparing the cluster X-ray temperature function
 with that observed at the present epoch.
Consistency with the estimated cluster abundance can be achieved for
COBE normalized MLM models with $\Omega_H/\Omega_M\le0.2$ and
  $ 0.45 \le \Omega_M \le 0.75$ for $h=0.5$.
  If $h=0.7$ then $ 0.3 \le \Omega_M \le 0.5$.
 These constraints are at $1\sigma$ level and standard MDM models are clearly 
 ruled out. 
  
We note that the range of allowed values for $\Omega_M$, that we obtain for MLM
models from linear analysis, is also approximately the same range that is
 needed in order to consistently satisfy a variety of independent 
 observational constraints. 
\end{abstract}

\section{Introduction}

In the standard framework of gravitational instability theory
present day structures must have been formed through the growth of small
inhomogeneities from an initial random Gaussian density field
 present at very early epochs,
with a scale-invariant Harrison-Zel'dovich spectrum.
Clustering analysis of the large scale structure in the
Universe has been improved in recent years by observations of
the spatial distribution of galaxies and cluster of
galaxies, as well as cosmic microwave background (CMB)
 anisotropies.
Thus any theory that wants to fit the observed large scale clustering must
be consistent with a set of constraints over more than three decades
in length: from galaxy correlation
($\sim 1 {h}^{-1} Mpc$, $H_0=100h Km sec^{-1}
Mpc^{-1}$ ) up to the
quadrupole CMB anisotropies detected by
COBE ($\sim 3000 {h}^{-1} Mpc$,
\cite{smo92}).

In the standard FRW  metric the fundamental
background cosmological parameters are related by a mutual relation.
It is also understood that for these parameters (the present matter density
 $\Omega_M$, the value of the
Hubble constant $H_0$
 and the age of the universe $t_0$)  the range of values allowed by
observations must be consistent with their FRW relation.
If we adopt the standard inflation
theory we shall assume that the total energy density in
the Universe is equal to the critical density ($\Omega_0=1$). On the
other hand, observations and theoretical predictions
suggest that the amount
 of baryon density must be small
($\Omega_b \sim 0.05 {(h/0.5)}^{-2}$, \cite{wal91}, \cite{cop95}).
A much larger contribution to the matter density must be of non-baryonic
nature. This is so-called dark matter (DM) problem and it has observational
support from dynamical estimates ($\Omega_M \sim {0.2 - 0.3}$,
\cite{bah96} and
references cited therein). It must be stressed that
there is not yet a firm evidence for $\Omega_M=1$, this
value for $\Omega_M$ being required by the inflationary paradigm.
The simplest possible model of DM is the one in which the universe is
 dynamically dominated by a single massive collisionless particle.
The most popular DM candidates are collisionless massive particles,
which decoupled from cosmological plasma either when being relativistic
(DM particles like massive neutrinos -- Hot DM) or non-relativistic
(hypothetical massive particles -- Cold DM).

The simplest model is CDM, where the power spectrum of the perturbations
depends on a single parameter,
 the present cold density $\Omega_C$.
Historically the first model to be considered was a neutrino
with a non-zero rest mass HDM (\cite{bon83}; \cite{zak80}).
The HDM model was soon rejected because numerical simulations ( \cite{whi83})
 produced nonlinear structures too late to be in agreement with
QSO existence.
The standard CDM model (SCDM) has been
analysed in considerable details
(\cite{dav85}; \cite {dav88}  and references cited therein).
It can match galaxy clustering ($\mincir 10 {h}^{-1} Mpc$) with a bias
parameter $b_g \magcir 2$, although it
lacks of sufficient power on large scales.
The SCDM model however became seriously inconsistent
with clustering data after the COBE detection
of a quadrupole anisotropy in the CMB (\cite{smo92}). When the
power spectrum of the model is normalized to the COBE
data SCDM models have unavoidable difficulties
(\cite{wha93}; \cite{oli93}; \cite{moa93}; 
Jing \& Valdarnini 1993; Dalton et al. 1991 ; Baugh \& Efstathiou 1994).

The Mixed
Dark Matter models (MDM) was  at first proposed~and~discussed~in a
few papers (Fang, Xiang \& Li 1984; Shafi \& Stecker 1984; 
Achilli, Occhionero \& Scaramella 1985; Valdarnini \& Bonometto 1985)
as an example to overcome the standard
Hot Dark Matter (SHDM).
Later, SMDM models were readressed again (\cite{hol89}; \cite{van92}).
Finally, it was proposed that a
certain mixture proportion of about $30/70\%$
for the Hot/Cold DM (\cite{sch92}; \cite{dav92}; \cite{tay92}; 
Klypin et al. 1993; Pogosyan \& Starobinsky 1993) 
as a  DM
model satisfying a wide body of observational data for clustering
( distributions of galaxies, galaxy clusters, quasars, Ly-$\alpha$ systems, etc.)
on typical scales
$\leq  l_{LS} \sim 100-150 {h}^{-1}Mpc$ and with an
Harrison-Zeldovich spectrum
of the primordial cosmological perturbations on scales $>l_{LS}$.
Because of the changes introduced into the power spectrum by
neutrino clustering, for SMDM models it was shown
the possibility
to reconcile the evidence of high coherent
velocity flows on large scales ($\sim 50 {h}^{-1}Mpc$ )
with the
moderate galactic pair velocities on megaparsec scales (\cite{kly93}).
For the cluster correlation function  ( \cite{hol93}; \cite{jin93};
\cite{klr93}) the two-point function for SMDM models
 appears to be consistent with data  for
 $R \magcir 20-30 h^{-1} Mpc$ ( on the contrary to what found for CDM ).

Analytical approximations for the present epoch transfer functions
are given by Holtzmann (1989);  Pogosyan \& Starobinsky (1995); Ma (1996);
Eisenstein \& Hu (1997).
Early numerical simulations have considered $\Omega_H=0.3$, we will use the
notation $H$ for massive neutrinos,
 but this model does not produce Ly-$\alpha$ systems as much as observed
(\cite{moa94}). For this reason a value of
$\Omega_H \mincir 0.2$ results in a better fit (\cite{kly95}; \cite{maa94}).
In comparison with SCDM or SHDM, MDM models are more complicated,
we have the choice of two independent parameters : the ratio
$\Omega_H/\Omega_C$ and $m_H$ ( if one allows for more than one species
of massive neutrinos). The spectral index $n$
of the post-inflationary spectrum is taken $n=1$.
It is possible to consider also the role of gravitational waves, which can
change significally
the normalization of the spectrum and the formation of structures
(e.g.,  \cite{maa96} ).

It must be stressed that SMDM models are in difficulty with the
present upper limits  on the age of the Universe:
if we assume $h=0.5$ and the age of globular
clusters greater than $15 Gyr$, in the case of a flat FRW  model the needed
value of $\Omega_M $ is $\leq 0.6 $.
The other difficulties are connected with later
galaxy and quasar formation
(\cite{pog95}; \cite{cen94}).
Another difficulty for SMDM models is that when the power spectra are
normalized to the COBE 4-yr data the linear theory overpredicits
the observed cluster abundances.
The estimated uncertainties in the normalization and linear calculations
can hardly fit the cluster number density within present data error bars.
The difficulty can be reduced if one removes the constraint of
a scale invariant spectra and introduces a small tilt ( $ n \simeq 0.8 -0.9$,
 \cite{maa96}).

In alternative to MDM models spatially flat low density
models with a positive cosmological constant
are considered to be a viable generalization of
SCDM after COBE. These models are
termed $\Lambda$CDM
and were considered even before the quadrupole detection
(\cite{pee84} and references therein). There are two main reason to consider
$\Lambda$CDM models as an alternative to MDM: the present
lower limit on the age of the Universe and the
baryon fraction in galaxy clusters.

Current uncertainties for the age of globular clusters give
$t_0=15 \pm 2 Gyr$ (\cite{cha96}). If one takes a lower
limits $t_0 \geq 13 Gyr$ then in the $\Omega_M=1$ case
the maximum value for $h$ is $0.5$.
This value is below the range allowed by recent
HST measurement : $h=0.7\pm.1$ (\cite{fre94}; \cite{rie95}).
The introduction of a positive cosmological constant allows $h$ to be higher
 for a fixed age $t_0$ than in $\Omega_M=1$.
The other reason to consider $\Lambda$CDM model is from X-ray
observations of galaxy clusters. If clusters are a representative sample of
the matter content in the Universe, then their baryons/total matter ratio
should not be different from what expected from standard nucleosynthesis
(\cite{whb93}). For $h \magcir 0.5 $ this is achieved
if $\Omega_M \mincir  0.35 \pm 0.2 $.
Quite interestingly this range is close to
that obtained from dynamical estimates. After COBE detection of the
quadrupole  anisotropy $\Lambda$CDM models have been further considered
(Kofman, Gnedin \& Bahcall 1993; Klypin, Primack \& Holtzman 1996; 
Liddle et al. 1996a; Liddle et al. 1996b).  
They~can fit several constraints on large scales
($k \mincir 1 {h}^{-1} {Mpc}^{-1} $, \cite{lib96}), but small scales
clustering it is excess (\cite{kly96}) by a factor $\approx 2-3$ when
compared to estimates from galaxy catalogs,  so this is a difficulty of the
model which can be solved with the introduction of an antibias of small
scales. A possibility which seems unplausible (\cite{pri96}).

 These difficulties have suggested that consistency with present data
 can be achieved for standard DM models with the introduction of
  one extra parameter. Possible variants are two species of
 massive neutrinos ( Primack 1997 ), or a tilt of the primordial
  spectrum ( Cole et al. 1997 ).

The main aim of this paper is to consider an
alternative possibility , that is MDM models with $n=1$ and
a non-zero cosmological constant (MLM).
These models have in fact the following free parameters :
$\Omega_{H}/\Omega_M$, $\Omega_\Lambda$ and the number of species of
massive neutrinos.
 If one considers one species of massive neutrinos, then MLM are on the
 same foot of the previous alternatives, the main advantage being
 that they retain the inflation paradigm and a scale-invariant
 initial spectrum.

In order to study these models we have numerically integrated forward in
 time the linearized Einstein-Boltzmann equations for the evolution of the
 metric and density perturbations.
The final power spectra have been normalized according to the
four year COBE data (\cite{bun97}).
We have then used linear perturbation theory
to find the parameters for which MLM models
are consistent with a variety of current observational constraints
on large scales (power spectrum, cluster-cluster correlations,
bulk velocities, cluster abundance).
In order to make comparison with observations we will consider the
following set of parameter space:
$\Omega_\Lambda=0, 0.31, 0.45, 0.65, 0.74$; $\Omega_H/\Omega_M=0.1, 0.2, 0.3$;
 number of massive and massless neutrinos $\beta_H=2, 4, 6$ and
$\beta_{\nu}=0, 2, 4, 6$ ( here $\beta$ is the sum of spin states ).
In some cases we considered a number of neutrinos species greater
than three, this possibility being of no physical relevance,
we ran these models
for studying  how different numbers  of massive and
massless particles  would change the spectra evolution.
Our final analysis will be restricted to one massive neutrino.
The fifth parameter is $h$ ($h=0.4, 0.5, 0.63, 0.7$).

The outline of the paper is as follows: in Section 2 we will present our model
description
and equations describing the linear perturbations in MLM models.
In Section 3 we will study the power spectra dependence on model
parameters using COBE 4-year data normalization.
In Section 4 some linear tests are given for MLM models, including
 our results for the cluster mass and temperature function.
Our conclusions are given in Section 5.

\section{Theoretical framework}
\subsection{Equations and method}
We have studied the evolution of density perturbations in a flat
 Friedmann cosmology.
In our models the present total density in critical units is
$\Omega_0=\Omega_C+\Omega_{H}+\Omega_{\Lambda}=1\equiv\Omega_M+
\Omega_{\Lambda}$.
 The notation $H$ means massive neutrinos, $C$ cold dark matter and
 $\Omega_{\Lambda} = \Lambda /(3 H_0^2)$.

 In our integrations we have followed numerically the time evolution
of the linearized Einstein-Boltzmann equations for the metric and adiabatic 
density
perturbations. We have considered scalar modes only and treated the
 perturbations for the following particle species : cold, massive and
massless neutrinos ($\nu$), photons.
We treat baryons and radiation as a single ideal fluid.
We have assumed that all collisionless particles
were decoupled from radiation before
the beginning of our computation.
In a collisionless medium the pressure anisotropy is different
from zero and the equations for density contrast and velocity are not
sufficient to describe the perturbations in $H$ and $\nu$ particles.
Collisionless components must be described by the
Boltzmann-Vlasov equation.
The density contrast, flux velocity and pressure anisotropy are given
by the moments of the perturbed distribution functions.

The set of equations for fluid and collisionless media,
with the Einstein equations for metric perturbations,
describes the evolution of density perturbations
in MLM models. The method of numerical solution of this system
is described in Valdarnini \& Bonometto (1985), and we refer to this paper for
more details.
The generalization of the equations to include a non-zero
cosmological constant is straightforward.

Our numerical calculation is
done for the following range of perturbation masses :
from $M=10^{20} M_{\odot}$ down to $M=10^{11} M_{\odot}$.
The integrations start
at the initial redshift $z_i=10^9$ and stop at the final epoch $z_f=5$.
 Because of the large region of parameter space that we have spanned with 
 our integrations, and the computational resources we had available, we
 have decided to stop the numerical integrations at a final epoch
 $ z_f > 0$. As a compromise between computing budget and 
 the accuracy  needed to evaluate observational linear variables 
 at the present epoch we chose $z_f=5.$

The initial conditions for our linear computations are given by
the Harrison-Zeldovich ( $n=1$ ) spectrum of density fluctuations:
$$
\mid \delta_C(k) \mid^2=A_i k  , \eqno(1)
$$
where $k$  is the comoving wavenumber of the perturbation and $A_i$ is an
arbitrary constant.
The total matter density perturbation is then defined as :
$$
{\delta}_M \equiv {{\delta \rho_C + \delta \rho_H} \over
{\rho_C + \rho_H}}={1 \over
{\Omega_M}}(\Omega_C \delta_C + \Omega_H \delta_H). \eqno(2)
$$

The transfer function $T(k)$ can be defined as the ratio of the amplitude of
the Fourier mode
$\delta_{M}(z_f,k)$ to the one of minimal $k$, which corresponds to the
wavenumber of the maximal mass perturbation $M=10^{20} M_{\odot}$.
The final linear transfer functions of our integrations  have then been 
 evaluated at $z=0$ using analytical formula.
The approximation involved neglects the changes in the shape
 of the transfer functions that take place between $z=5$ and $z=0$ because
 of the decrease in the neutrino streaming. 
 The error involved in the computation of the linear variables, with
  which we test the models against a set of data, is negligible
 in most of the considered cases and amounts to a few percent when
 $\Omega_H/\Omega_M=0.2$ and $\Omega_M =0.25 $.

To describe the evolution of fluctuations between $z_f$ and $z=0$ 
we have applied the well known exact analytical solutions of Einstein
equations for a
perturbed flat dust model with non-zero $\Lambda$ (\cite{kof85,lah91}).

A non-zero cosmological constant changes the expansion rate so that the scale
factor has the following expression
$$
a(t)=\left({\Omega_M \over
\Omega_{\Lambda}}\right)^{1/3} \sinh^{2/3}({3\over 2} \sqrt{\Omega_{\Lambda}}
H_{0}t), \eqno(3)
$$
and as a result the growth of perturbations relative
to that of a critical density universe, after
the moment of equality of the cosmological constant density to the
matter one, it is suppressed.
 The suppression coefficients for density and velocity perturbations
are the following:
$$K_{\delta}(t)={5\over 3}\left(1-{{\dot a}\over
a^{2}}\int_{0}^t a(t)dt\right),\eqno(4)$$
$$K_{v}(t)={5\over 3}\left({{\dot a}\over a^{2}}-{\ddot a\over a{\dot
a}}
\right)\int_{0}^t a(t)dt.\eqno(5)$$
The total suppression at z=0 is well approximated by
$$
g(\Omega_M)={5\over 2}\Omega_M\left[{1\over
70}+{209\Omega_M-{\Omega_M}^2
\over 140}
+{\Omega_M}^{4/7}\right]^{-1},\eqno(6)
$$
for density perturbations (\cite{car92})
and by ${\Omega_M}^{0.6}$ for peculiar velocities (\cite{lah91}).

We express the final matter power spectrum
which gives rise to the observed large scale structure of the
Universe as
$$
P(k)=AkT^{2}(k) K^{2}_{\delta}(t_{0}) /\Omega_M^{2}, \eqno(7)
$$
where $A$ is a normalization constant and $T(k\rightarrow0)\rightarrow 1$ is
 the total matter transfer function.
 In Eq.(7) $P(k)$ is the density weighted power spectrum, i.e.
$P(k)= \left [\Omega_C P_C^{1/2} +\Omega_H P_H^{1/2} \right ]^2$.

\subsection{Normalization procedure}

In order to constrain a cosmological model one has to fix the normalization
amplitude of the density fluctuation power spectrum at the present epoch.

The most accurate method for normalizing $P(k)$
is to make use of the COBE satellite data for
cosmic microwave background (CMB) anisotropies (\cite{smo92,ben94,ben96}).
Bunn \& White (1997) have carried out
a likelihood analysis of the 4-year COBE DMR sky maps and have obtained the
best-fitting quadrupole, the value of which depends strongly on the
initial index $n$, $\Omega_M$ and the space curvature. For example, for a
Sachs-Wolfe $n=1$ spectrum $<Q>=18.7\pm1.3 \mu K$, for $n=1.5$ $<Q>=13\pm0.8
\mu K$; in the case of flat cosmological models with non-zero $\Lambda$ and open models
its value increases when $\Omega_M$ decreases.
For the normalization of the spectra in different cosmologies
 Liddle \& Lyth (1993) have proposed instead of the quadrupole to use
 the amplitude of the
density perturbation at
 horizon-crossing $\delta_{hor}$ defined by
$$
\Delta^{2}(k)={k^{3}P(k)\over 2\pi^{2}}=\delta_{hor}^{2}\left({ck\over
H_{0}}\right)^{3+n}T^{2}(k),\eqno(8)
$$
which fixes the present-day normalization of spectra.
A fit to the 4-year COBE data for flat models with the $n=1$
post-inflation spectrum analysed here has the following simple form
(\cite{lia96}; \cite{lid93}; \cite{lib96}; \cite{bun97}):
$$
\delta_{hor}(\Omega_M)=1.94\;10^{-5}\Omega_{M}^{-0.785-0.05ln\Omega_M}.\eqno(9)
$$

Proceeding from the definitions of $\delta_{hor}$ and the power spectrum, 
Eq.(7),
the normalization constant is calculated as
$$
A=2\pi^{2}\delta_{hor}^{2}\left({c\over H_{0}}\right)^4\left({\Omega_M
\over K_{\delta}}\right)^2.\eqno(10)
$$

\section{Power spectra}

The power spectra of Eq.(7), normalized to the COBE 4-year data,
are depicted in Figs. 1-3 for different models.
\begin{figure}[!ht]
{\psfig{file=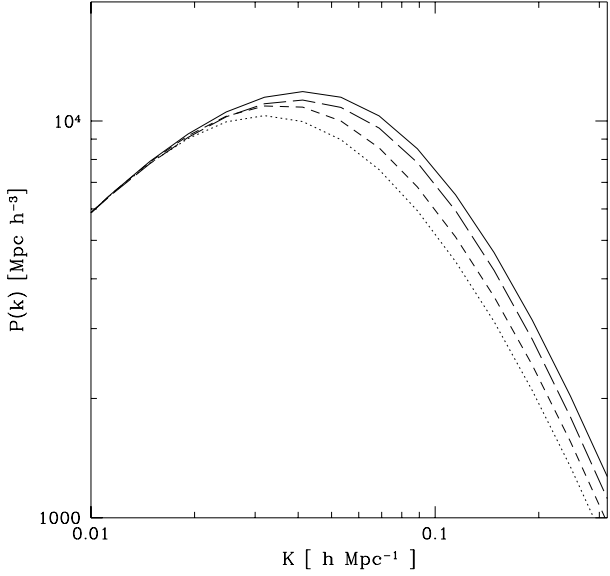,width=9.0cm}}
\caption[]{ The linear density power spectrum $P(k)$, normalized to the 
4--yr COBE data, is shown for models with $\Omega_{\Lambda}=0$,
$\Omega_H/\Omega_M=0.2$, $h=0.5$ and different
values of the $\beta_H$ and $\beta_\nu$ parameters.
The values chosen are:
$\beta_H=2 \ \beta_\nu=4$ (solid line),
$\beta_H=6 \ \beta_\nu=0$ (dotted line),
$\beta_H=4 \ \beta_\nu=2$ (short-dashed line),
$\beta_H=2 \ \beta_\nu=6$ (long dashed line)}
\label{fig1}
\end{figure}

Fig. 1 shows how the number of species of massive neutrinos,
when $\Omega_{H}$  is kept fixed ($=0.2$),
modifies the amplitude of the spectra at $k>0.1h Mpc^{-1}$.
As we can see, increasing the number of
 species of massive neutrinos from 1 to 3 decreases the power at
scale $< 100 h^{-1}Mpc$ of about 1.6 times. Increasing the number of species
of massless neutrinos when the number of the massive ones is fixed 
also suppresses the
power.
Both of these effects have a simply explanation.

The dynamics of the perturbations in the hot components depends on two
characteristic scales: one of them is
the free-streaming scale, which is about the inverse of the Hubble
scale, when the
particles first become non-relativistic.
In our case
$k_{H}=k_{H}(z_{H})\simeq H(z_{H})/c \simeq {m_H\over200}\;Mpc^{-1}$
where
$z_{H}$ is the redshift at which the $H$-particles become non-relativistic,
 $m_{H}$ is in $eV$ and $H(z)$ is the Hubble function.
\begin{figure}
{\psfig{file=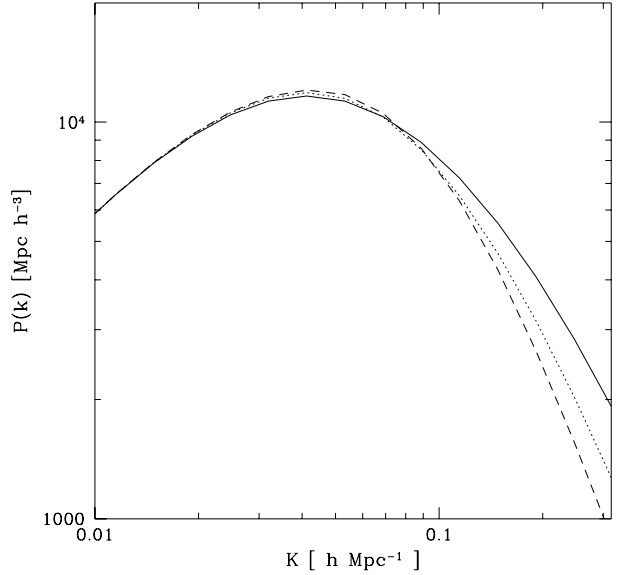,width=9.0cm}}
\caption[]{We show the power spectra dependence on the $\Omega_H/\Omega_M$
parameter. The models chosen are MDM models with
$\Omega_{\Lambda}=0$, $h=0.5$,
$\beta_H=2, \beta_\nu=4$ and different values of $\Omega_H$ :
$\Omega_H=0.1$ (solid line),
$\Omega_H=0.2$ (dotted line) and $\Omega_H=0.3$ (short-dashed line)}
\label{fig2}
\end{figure}

The second one is connected to the
free-streaming scale, i.e.
 the minimal size $ \simeq k_H(z)^{-1}$ of  $H$-particle objects that
can collapse at a given redshift $z$.
After the epoch $z_H$ the velocities of massive neutrinos are redshifted away
adiabatically, then $k_{H}^{-1}(z)$ decreases.
On large scales, $k<k_{H}$, gravitational instability  develops
in a standard way. For the growing mode of
adiabatic perturbations $\delta \simeq \delta_{H}\simeq \delta_{C}\sim a$.
On the other hand the growth of density perturbations on smaller scales,
$k>k_{H}$,
is reduced from the moment when a perturbation of a given scale enters
the horizon and up to the time when the free-streaming scale becomes
smaller than the perturbation size.
Because of the free-streaming effects density perturbations
are erased away in the part of $\sim\Omega_{H}$, they still grow in
the cold component in such a way that
the total density growth, although less than in CDM, appears much
larger than in HDM model. Finally for $k<k_{H}(z)$ the perturbations
in both components develop similarly: the dynamical amplitude
$\delta_{H}$ gradually approaches the other component $\delta_C$.
\begin{figure}
{\psfig{file=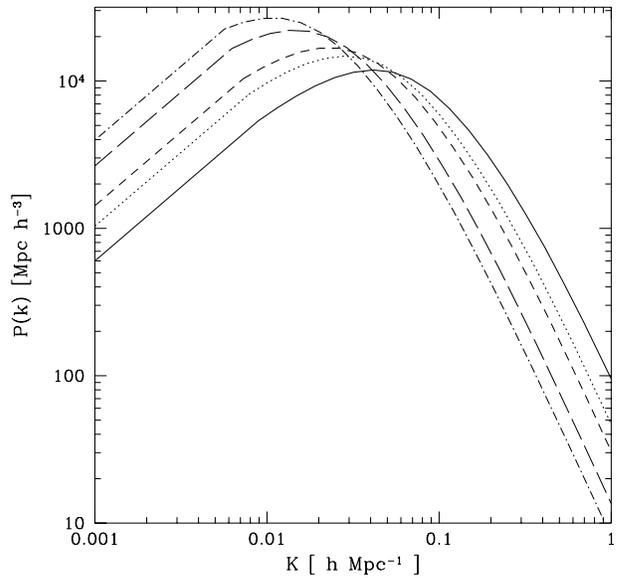,width=9.0cm}}
\caption[]{Power spectra are shown for MLM models
with $h=0.5$,
$\beta_H=2, \beta_\nu=4$, $\Omega_H/\Omega_M=0.2$ and different
values of the $\Omega_\Lambda$ parameter:
$\Omega_\Lambda=0.$ (solid line),
$\Omega_\Lambda=0.31$ (dotted line),
$\Omega_\Lambda=0.45$ (short dashed line),
$\Omega_\Lambda=0.65$ (long dashed line),
$\Omega_\Lambda=0.74$ (dot-short dashed line)}
\label{fig3}
\end{figure}

Thus decreasing the mass of massive
neutrinos (increasing their number of species when $\Omega_{H}$ is fixed)
shifts to later epochs the time when they become non-relativistic and the
collisionless damping becomes more effective in slowing down the growth of 
matter density perturbations.

Increasing the number of species ($N_{\nu}$) of massless neutrinos
results in a longer duration of the radiation dominated era
and in a smaller redshift
$a_{eq}^{-1}=1.+z_{eq} \simeq 4.16 \, 10^4 \Omega_M h^2 / (1+0.227 N_{\nu})$
of the matter-radiation equality.
It is useful to define the comoving wavenumber $k$ with respect to the scale
that crosses the horizon at the matter-radiation epoch, i.e.
$$
q= k/k_{eq}= k / [ 0.4 \, \Omega_M h^2(1+0.227N_{\nu})^{-1} Mpc^{-1}].
$$
Because of the approximate scaling of the transfer function with the
dimensionless parameter $q$,  an increase in $N_{\nu}$ implies a reduced
amplitude of the power spectrum  at a given $k$.

So both of these effects, increasing the number of massive and massless
neutrinos, increase the duration of the
epoch when the growth of density perturbations is less effective.
\begin{figure}
{\psfig{file=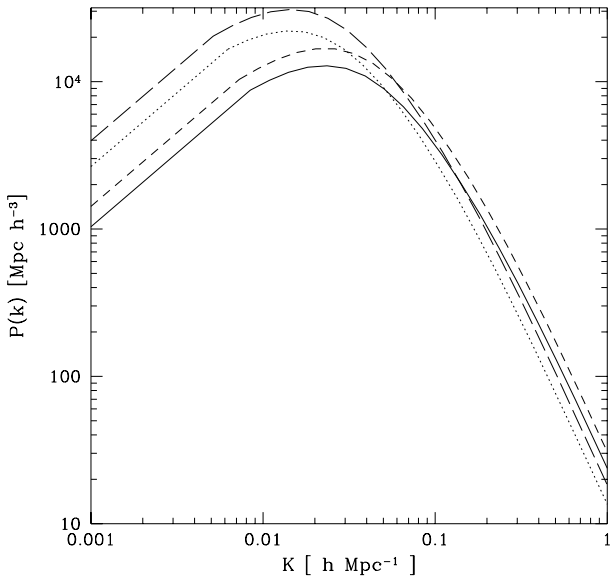,width=9.0cm}}
\caption[]{Here $P(k)$ is plotted for two groups of MLM models with fixed
parameter
$\Omega_M h$, $\beta_H=2, \beta_\nu=4$, $\Omega_H/\Omega_M=0.2$.
For $\Omega_M h=0.275$:
$\Omega_M=0.69 \: h=0.4$ (solid line),
$\Omega_M=0.55 \: h=0.5$ (dashed line).
For $\Omega_M h=0.175$:
$\Omega_M=0.35 \: h=0.5$ (dotted line),
$\Omega_M=0.26 \: h=0.67$ (dot dashed line)}
\label{fig4}
\end{figure}
The dependence of the spectra on $\Omega_{H}$, or on the neutrino rest mass,
for one massive and two massless neutrino species is shown in Fig. 2.
Increasing $\Omega_{H}$ from 0.1 to 0.3 decreases the power at
galaxy scales by a factor of about $ \sim 4$ , at galaxy cluster scales  by $\sim1.4$ . The inverse behavior
of these spectra near the maximum at $k\simeq0.04-0.05h Mpc^{-1}$ is caused by
 the existence of three characteristic timescales for the hot component:
 the epoch when it becomes non-relativistic,
the period of collisionless damping and  when  the energy density of massless
neutrinos equals the massive one.

The power spectra for different $\Omega_{\Lambda}$ and fixed
values for the parameters $\beta_H, \beta_{\nu} ,
\Omega_H/\Omega_{M}$ are shown in Fig. 3.
The horizontal shift of the spectra is caused
by the later epoch of equality of the energy densities for the  $\nu$ and $H$
 particles.
In different $\Lambda$ models with fixed $\Omega_H$ the epoch of equality is
different, in particular increasing the value of $\Omega_\Lambda$ then
$\Omega_C$ decreases and $z_{eq}$ decreases too.
The vertical shift is caused by the coefficient $K_{\delta}/\Omega_M$ in Eq.(10)
and the amplification of CMB anisotropies in non-zero $\Lambda$ models,
which is included in the normalization parameter $\delta_{hor}$.
The change of $\Omega_{H}$ in non-zero $\Omega_\Lambda$ models changes the
spectra in a way similar  to the $\Omega_\Lambda=0$ case.

\begin{figure*}[!ht]
\centerline{\hbox{%
\psfig{figure=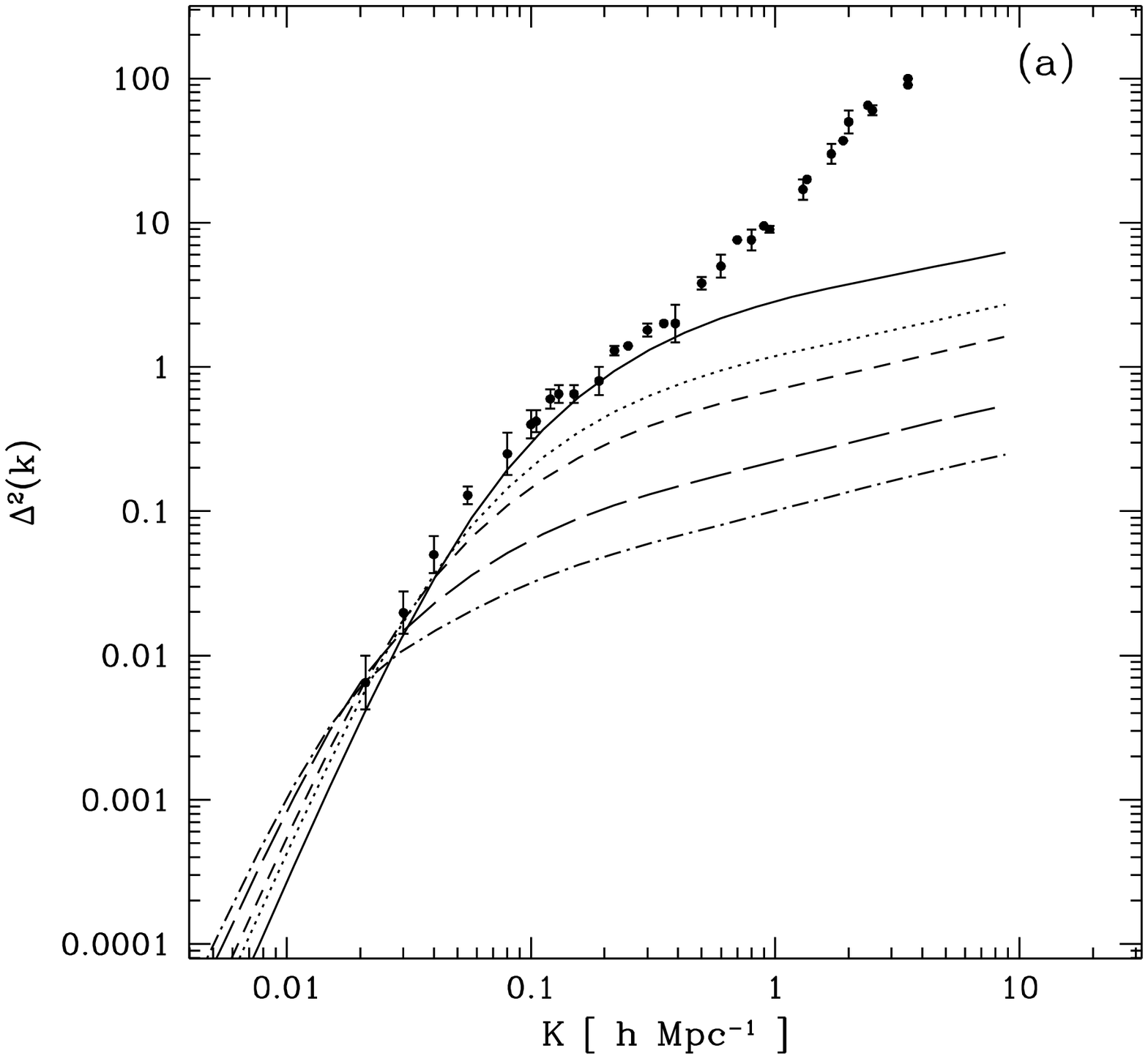,height=8truecm,width=8truecm}%
\psfig{figure=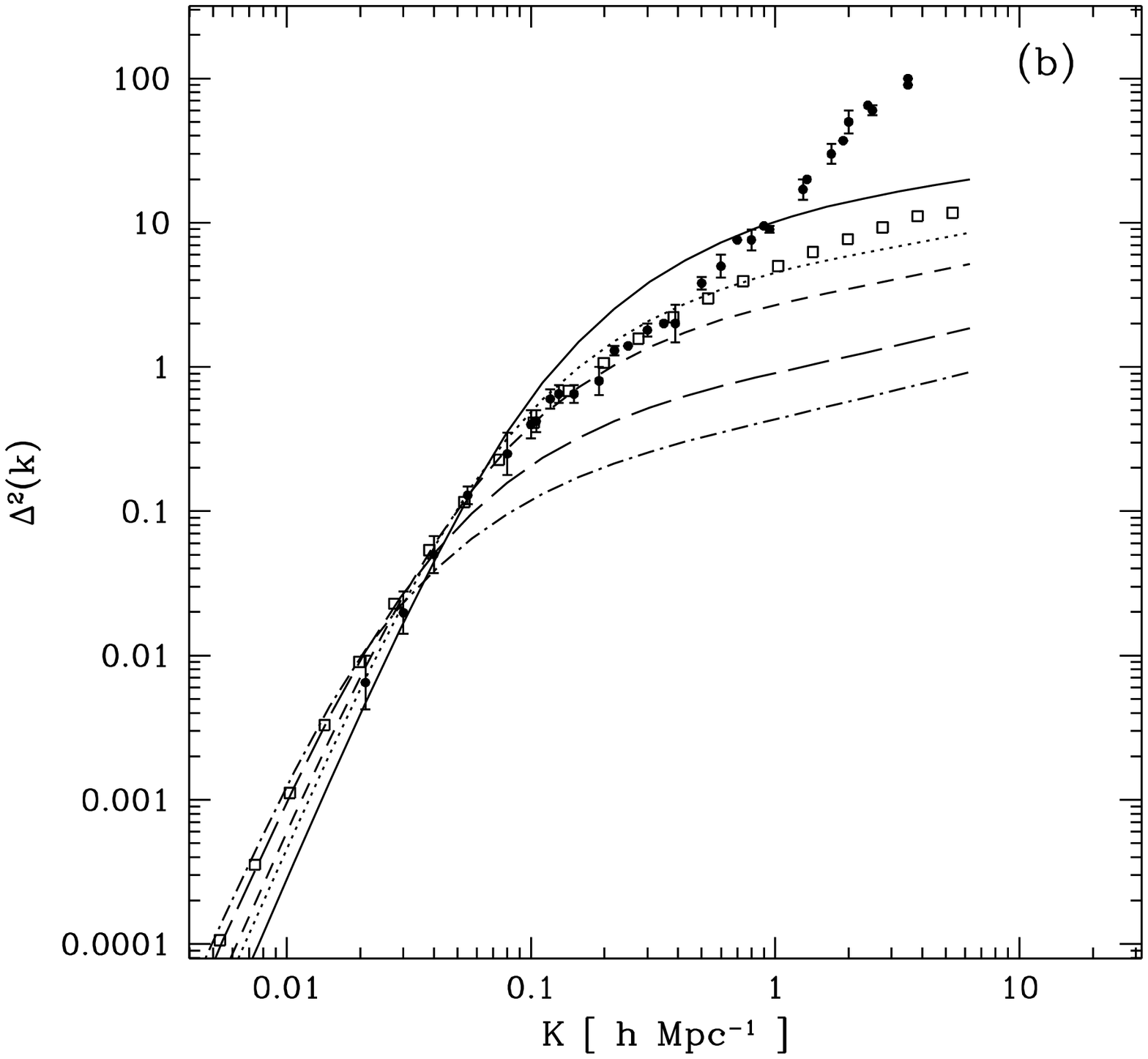,height=8truecm,width=8truecm}%
}}
\caption[]
{{\bf a} $\Delta^2(k)$ is shown for MLM models with $h=0.5$,
$\beta_H=2, \beta_\nu=4$, $\Omega_H/\Omega_M=0.2$ and different
values of the $\Omega_\Lambda$ parameter:
$\Omega_\Lambda=0.$ (solid line),
$\Omega_\Lambda=0.31$ (dotted line),
$\Omega_\Lambda=0.45$ (short dashed line),
$\Omega_\Lambda=0.65$ (long dashed line),
$\Omega_\Lambda=0.74$ (dot-short dashed line).
Black circles are the real space APM power spectrum
obtained by Peacock (1997). {\bf b} The same as in panel {\bf a}, but for
 $h=0.7$. Open squares represent the standard $\Lambda$CDM model with
 $\oml=0.7$, normalized to $\sigma_8=1$}
\label{fig5}
\end{figure*}

The following property of spectra
is suggested from our calculations: the models with the same component
species ($\beta_{H},\beta_{\nu},\Omega_{H}/\Omega_{M}$) but
different
$\Omega_M$ and $h$ have the same form and localization of the maximum of
$P(k)$
when $\Omega_M h$ 
is constant. This is shown in Fig. 4 for two values of
$\Omega_M h$: 0.176 and 0.275.

\section{Large scale structure: predictions versus observations}
\subsection{Power spectra data and models}

 In order to test our models we compare the linearly evolved
  $P(k)$ with the one obtained from clustering data.
Our power spectra $P(k)$ have been computed for MLM models with a zero-baryonic
content, in order to properly compare with data we then correct the shape
 of the transfer functions according to a baryonic fraction 
 $\Omega_b =0.015/h^2$ (see sect.4.4).

We will make use of the dimensionless spectrum
$\Delta^{2}(k)=k^{3}P(k)/2\pi^2$, calculated at $z=0$, and compare it
 with the real space power spectrum of optical and IRAS galaxies
 obtained by  Peacock (1997) and Peacock \& Dodds (1994), taking
 into account redshift distortions and bias.
 Because we have performed a linear calculation we
 restrict ourselves to comparing $\Delta^{2}(k)$ only in its linear
 part $k \mincir 0.2 hMpc^{-1}$.
 For a linearly evolved spectrum the deviation from non-linearity
 because of gravitational clustering
 becomes significant  at $k \magcir 0.5 hMpc^{-1}$.

In Figs. 5 \& 6 we show $\Delta^2(k)$  for a set of
MLM models with different values of  $\Omega_{\Lambda}$
and $\Omega_H/\Omega_M=0.2$ (Fig. 5),
$\Omega_H/\Omega_M=0.1$ (Fig. 6). We assume one species of massive neutrinos.

 For each case we show two subcases: $h=0.5$ and $h=0.7$.
 For $h=0.7$ we plot also the
 $\Delta^2(k)$ of the $\Lambda$CDM
 model with $\Omega_{\Lambda}=0.7$, normalized to an rms mass
fluctuation $\sigma(R=8h^{-1}Mpc)\equiv \sigma_8 =1$ (see Eq.(19)).
 For this model we will use the notation $\Delta^2_{\Lambda}(k)$
 for $\Delta^2(k)$.
 The observational data are shown as black circles and
 correspond to the real space APM power spectrum for galaxy data
(Peacock 1997).
We make use of these data, rather than those of 
Peacock \& Dodds (1994), because of the improved treatment 
for clustering evolution performed in the former paper.

 In order to constrain the parameter space for different models we
 have preferred not to do
 a $\chi^2$ analysis because of the correlation between different bins.
Also because of the statistical uncertainties we consider
it more appropriate to make a qualitative comparison.
We formally define a given model to be consistent with clustering data 
if the computed $\Delta^2(k)$ are within the $2 \sigma$ errors for
$k \mincir 0.1 hMpc^{-1}$.

From Fig. 5a the model predictions for $\Delta^2(k)$ in the case of
large values of $\Omega_{\Lambda}$ ($\magcir 0.3$)
are below  the observational data in the linear $k$ region.

 This limit on $\Omega_{\Lambda}$  can be reduced if one considers
 $h=0.7$ (Fig. 5b). In this case the inconsistency in the linear regime is for
 $\Omega_{\Lambda}$ $\magcir 0.65$.
 This shift in the spectrum follows from the dependencies discussed
 in the past section due to the increase in $h$.

 It is important to note the drastic reduction of small-scale power
 in MLM models because of massive neutrinos.
  In the parameter space the closest model to $\Lambda$CDM, for the
  set of MLM models of Fig. 5b, is $\Omega_{\Lambda}= 0.65$.
 At high wave-number the spectrum of this model is well below the
 $\Delta^2_{\Lambda}(k)$ of the $\Lambda$CDM.
For ease of comparison the $\Lambda$CDM has
 been normalized to $\sigma_8=1$, while the COBE normalization yields
 $\sigma_8 \simeq 1.1$. On the other hand, for the MLM models considered,
  we always found $\sigma_8 \mincir 1$ (see sect.4.5).
We thus conclude that MLM models  have the property of
removing the unpleasant feature of galaxy anti-biasing, invoked for $\Lambda$CDM
 (\cite{kly96}) in order to fit clustering data.
\begin{figure*}[!ht]
\centerline{\hbox{%
\psfig{figure=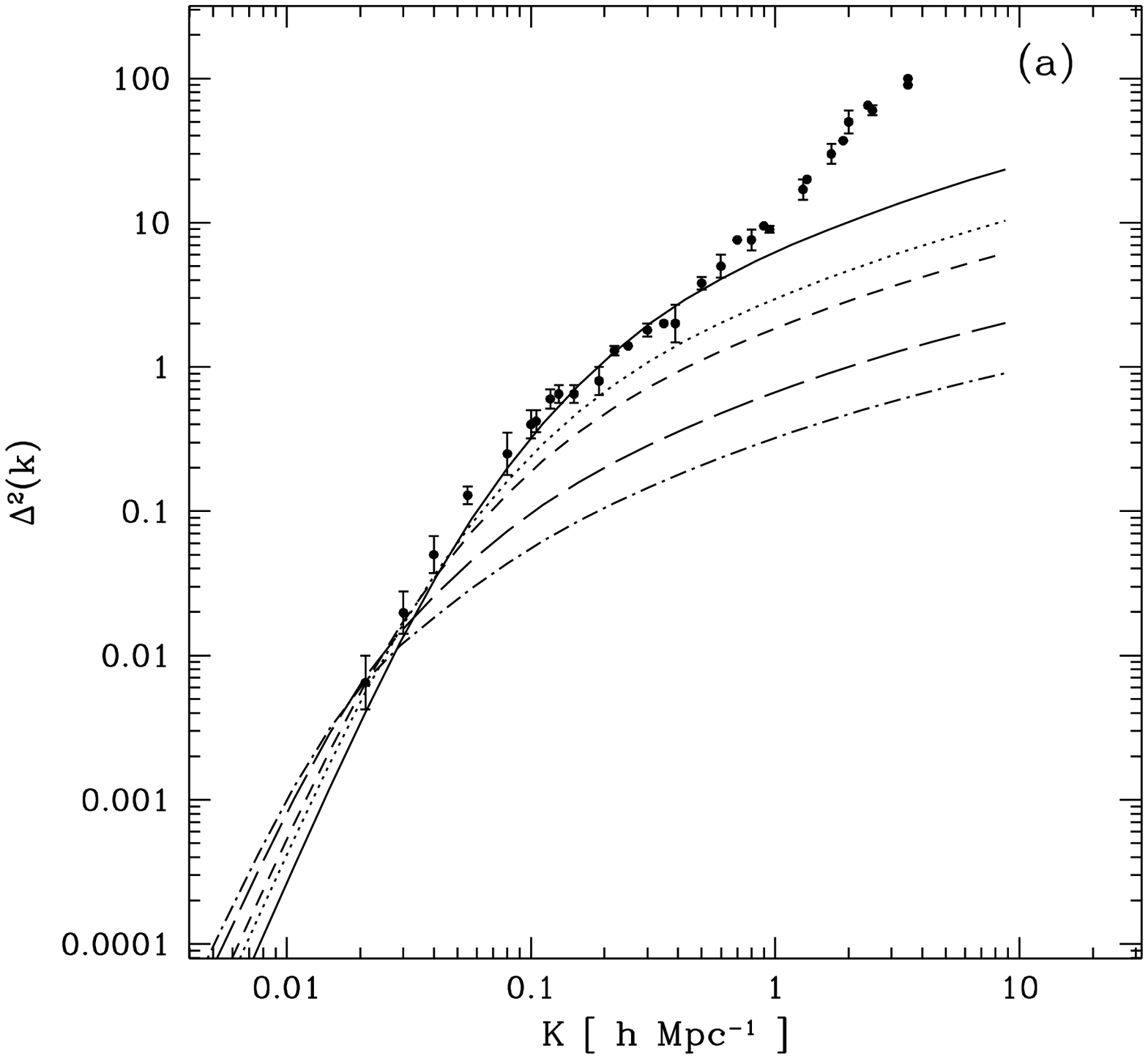,height=8truecm,width=8truecm}%
\psfig{figure=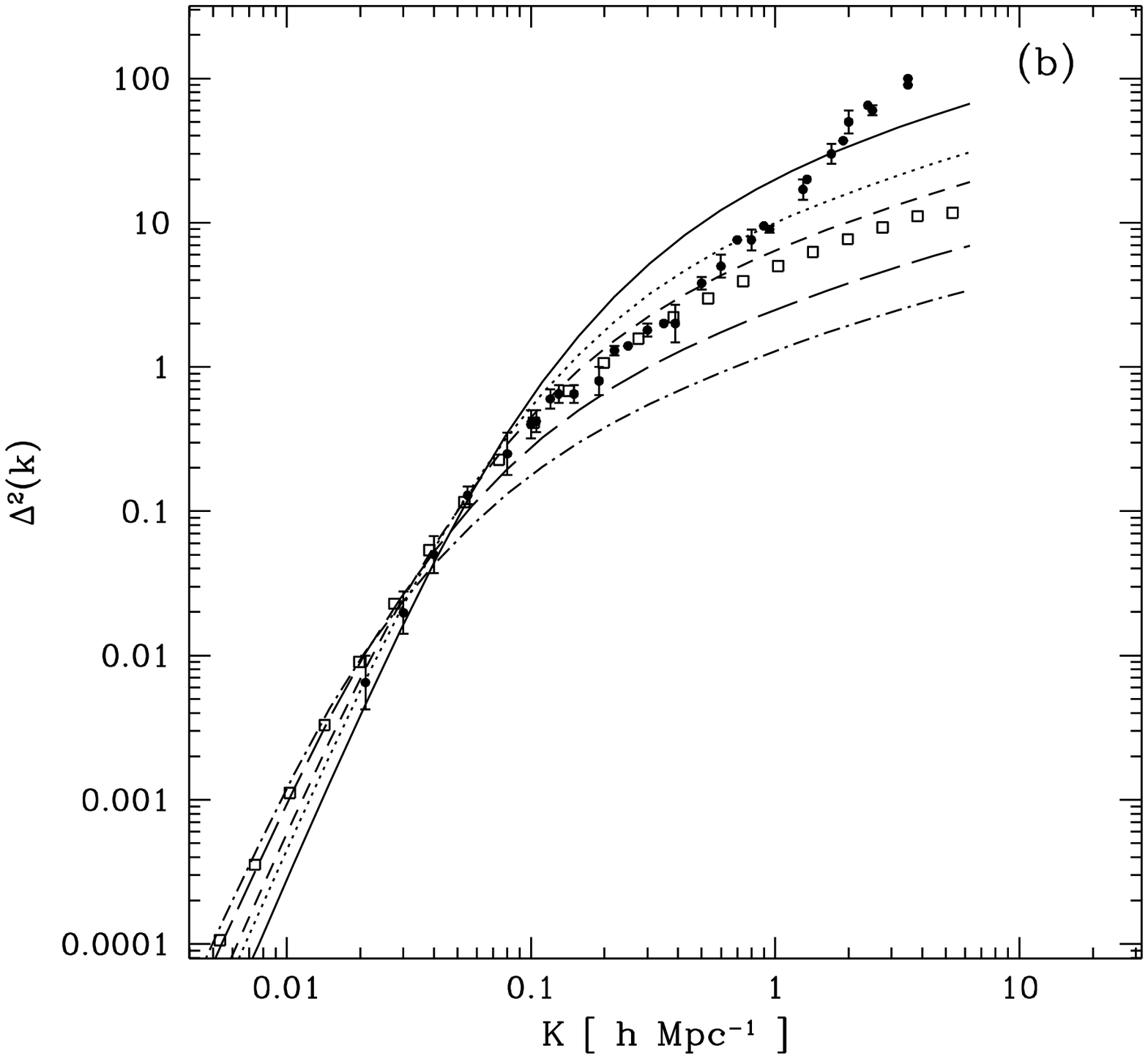,height=8truecm,width=8truecm}%
}}
\caption[]
{{\bf a} $\Delta^2(k)$ are shown for the same set of MLM models 
of Fig.5a, but with a ratio $\Omega_H/\Omega_M=0.1$.
{\bf b} The same as in panel {\bf a}, but for $h=0.7$}
\label{fig6}
\end{figure*}

As we can see from Fig. 5 the spectra of MLM models
with $\Omega_{H}/\Omega_M  =0.2 $ do not contradict on linear scales
 the spectrum reconstructed from observations
for $\Omega_M  \magcir 0.7 (0.35)$ and $h=0.5 (0.7)$.
  Fig. 6 refers to $\Omega_{H}/\Omega_M  =0.1 $  and consistency is
  achieved for $\Omega_M  \magcir 0.55 (0.25)$ and $h=0.5 (0.7)$.

 This range of limits can be further constrained if one considers that
 non-linearity effects due to gravitational clustering will enhance
 $\Delta^2(k)$ well above our linear $\Delta^2(k)$ , for
 $k \magcir 0.2 hMpc^{-1}$ (Peacock 1997). We then require that our
 linear spectra should be at least below the observed reconstructed spectrum
  in the high-$k$ region.
 Furthermore, for $h=0.7$, we already know that the $\Lambda$CDM over produces
 small-scale power, then the $\Delta^2(k)$ of our MLM models
 should not exceed  the $\Delta^2_{\Lambda}(k)$ at high wave-numbers.

The case $h=0.5$ does not give useful constraints, while for $h=0.7$
we obtain  $0.35 \mincir \Omega_{M} \mincir 0.55 $ for
$\Omega_H / \Omega_M =0.2$; $0.25 \mincir \Omega_{M} \mincir 0.35 $ for
$\Omega_H / \Omega_M =0.1$.
The constraints that we obtain are for a constant linear bias, with
$b_{gal} =1$. 

If the optical data are a biased tracer of the dark matter 
then these limits should be changed according to the value of $b_{gal}$. 
However no simple scaling is possible for the  obtained constraints
because the value of $b_{gal}$ will change for any model according 
to the value of $ \Omega_M $  and $\Omega_H / \Omega_M $.
In sect.4.5 we compare the constraints on $\Omega_M $ that we obtain from
cluster abundances  with those given by clustering data , according to
the value of $\sigma_8$.

We do not attach a particular statistical significance to
these lower limits on $\Omega_M$, nevertheless in some cases we found 
these limits to be 
consistent with the constraints on $\Omega_M$ obtained
independently from cluster abundances (see sect.4.5).

\subsection{Bulk motions}
Another constraint on dark matter models comes from the study of galaxy
bulk flows in spheres around our position (see  e.g., \cite{kof93};
\cite{sto95}; \cite{lib96}).
Bertschinger et al. (1990) and Dekel (1994) give
the average peculiar velocities within spheres of radius between 10 to
60 $h^{-1}Mpc$ after previously smoothing raw data with a Gaussian
filter of radius $R_{f}=12h^{-1}Mpc$. When the power spectrum is known then
the rms peculiar velocity of galaxies in a sphere of radius $R$,
corresponding to these data, can be calculated with the following
expression
$$
V^{2}(R)=H^{2}_{0}K^{2}_{v}K^{-2}_{\delta}/2\pi^{2}
\int_{0}^{\infty}
P(k)exp(-k^{2}R_{f}^{2})W^{2}(kR)dk,\eqno(11)
$$
where $W(kR)$ is the top-hat window function.

The calculated predictions for the rms bulk motions for different MLM models
with $h=0.5$ are shown in Fig. 7.
As we can see, decreasing $\Omega_M$ effectively reduces the
bulk motion and the explanation of the observed data in low $\Omega_M$ models
is problematic. So, in models with  $\Omega_M<0.3$ and
$\Omega_{H}/\Omega_M\ge0.2$
the observable data are above the  $95\%$ confidence level of the rms
predictions. A similar conclusion holds if $h=0.7$.
But taking into account the large error bars we must
admit that these data do not rule out any of the models analysed here.
We only conclude that models with $\Omega_M>0.3$ and
$\Omega_{H}/\Omega_M\le0.2$ are preferred.

\begin{figure}
{\psfig{file=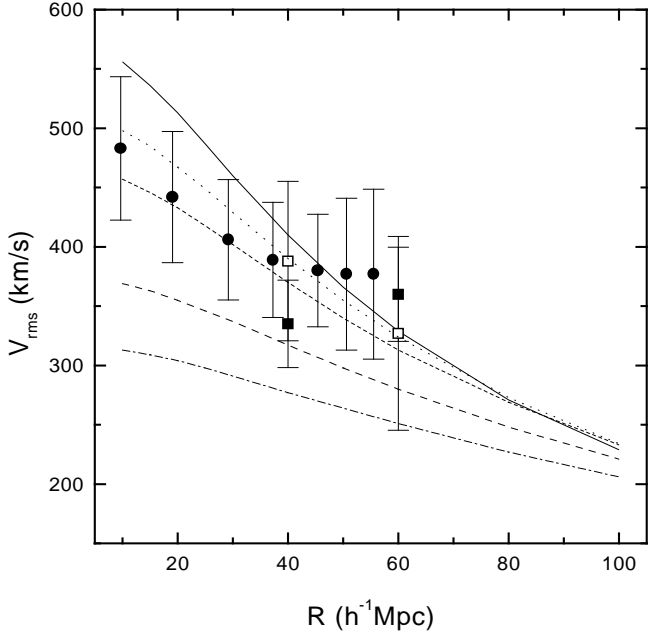,height=9.0cm,width=9.0cm}}
\caption[]
{The bulk motion
$V_{rms}(R)$ is shown for MLM models with $h=0.5$,
$\beta_H=2, \beta_\nu=4$, $\Omega_H/\Omega_M=0.2$ and different
values of the $\Omega_{\Lambda}$ parameter:
$\Omega_{\Lambda}=0$ (solid line),
$\Omega_{\Lambda}=0.31$ (dotted line),
$\Omega_{\Lambda}=0.45$ (short dashed line),
$\Omega_{\Lambda}=0.65$ (long dashed line),
$\Omega_{\Lambda}=0.74$ (dot-short dashed line).
The solid circles correspond to Dekel(1994),
open squares to Bertschinger et al.(1990), solid squares to Courteau et al.
(1993) }
\label{fig7}
\end{figure}

\subsection{Cluster-cluster correlations}

The space distribution of rich clusters is a powerful tool for probing the
power spectrum at intermediate and large scales.
The first attempts  to measure the two-point spatial autocorrelation
function $\xi_{cc}(r)$ (\cite{bah83}, \cite{kly83}) have shown
that they are more clustered than galaxies and $\xi_{cc}(r)\approx
(r_{0}/r)^{\gamma}$ with $\gamma=1.6-2.0$ and $r_{0}=16-25\;h^{-1}Mpc$.

\begin{figure}
{\psfig{file=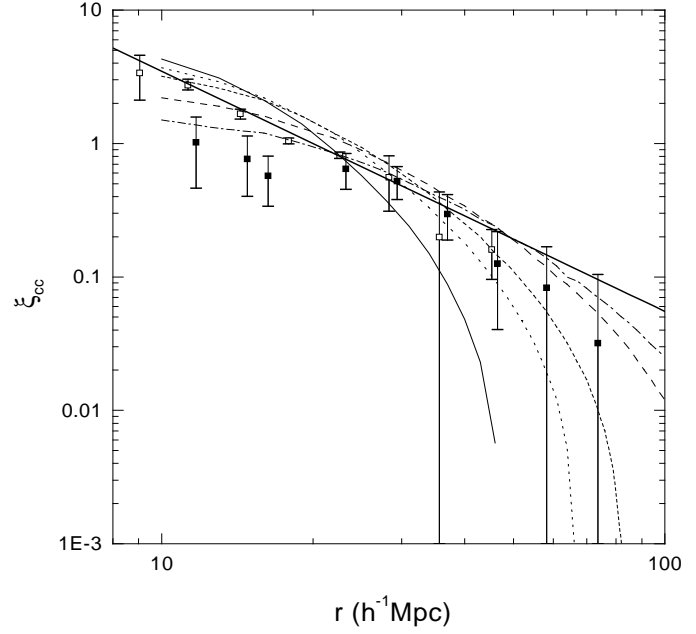,height=9.0cm,width=9.0cm}}
\caption[] {The cluster-cluster correlation functions
$\xi_{cc}$ are shown for
MLM models with $h=0.5$, $\beta_H=2, \beta_\nu=4$,
$\Omega_H/\Omega_M=0.1$ and different
values of the $\Omega_{\Lambda}$ parameter:
$\Omega_{\Lambda}=0$ (solid line),
$\Omega_{\Lambda}=0.31$ (dotted line),
$\Omega_{\Lambda}=0.45$ (short dashed line),
$\Omega_{\Lambda}=0.65$ (long dashed line),
$\Omega_{\Lambda}=0.74$ (dot-short dashed line)  }
\label{fig8}
\end{figure}

The later analysis of other authors (see for example Postman et al. 1992;
Olivier et al. 1993; Jing \& Valdarnini 1993)
has confirmed that $\xi_{cc}$ is well fitted by the same
expression with $\gamma=1.8$ and $r_{0}\approx 20\;h^{-1}Mpc$. The
important conclusion, which follows from numerous studies of this
problem, is the existence of a positive long distance correlation of
rich clusters of galaxies out  to $50\;h^{-1}Mpc$.
For a Gaussian random density fluctuation field the
correlation function of peaks at large separations is
calculated with the following equation (\cite{bar86}):
$$
\xi_{cc}(r)={b_{c}^{2}\over 2\pi^{2}}\int_{0}^{\infty}
P(k)k^{2}W^{2}(kR_{c}){sin\;kr\over kr}dk,\eqno(12)
$$
where $W(kR_{c})$ is the window function, which filters out in the density
field
the structures on scales larger than $R_{c}$, $b_{c}$ is their biasing
parameter, which takes into account the statistical correlation of peaks
above a given threshold. The biasing parameter
is defined by the expression
$$
b_{c}=<\tilde \nu>/\sigma(R_{c})+1,\eqno(13)
$$
where the effective threshold level $<\tilde \nu>$ is given by
$$
<\tilde \nu>=\int_{0}^{\infty}d\nu \left(\nu-{\gamma \theta\over
1-\gamma ^{2}}\right)t(\nu/\nu_{t})N_{pk}(\nu)/n_{c}\eqno(14)
$$
with $\gamma$, $\theta$, the threshold function $t(\nu/\nu_{t})$ and the
differential number density $N_{pk}(\nu)$ are defined according to
equations (4.6a), (6.14), (4.13) and (4.3)
of Bardeen et al. (1986).
The Gaussian filter radius
corresponding to the mass
($M\ge 5\;10^{14}h^{-1}M_{\odot}$)
of a rich cluster is
$R_{c}=(M/4.35\;10^{12}\Omega_M h^{-1})^{1/3}h^{-1}Mpc$.
The observed number density of Abell clusters with richness
$R\ge 1$, $n^{obs}_{c}=(5.7\pm 0.5)\;10^{-6}h^3 Mpc^{-3}$
(see \cite{zam91}; \cite{bah88} for a review), is used for the
determination of the peak height $\nu_{t}$

$$
n_{c}=\int_{0}^{\infty}d\nu t(\nu/\nu_{t})N_{pk}(\nu)=n^{obs}_{c}.\eqno(15)
$$

The mean height of peaks for which the rich clusters of galaxies are formed
is then
$$
<\nu>=\int_{0}^{\infty}d\nu t(\nu/\nu_{t})\nu N_{pk}(\nu)/n_{c}.\eqno(16)
$$

\begin{figure}
{\psfig{file=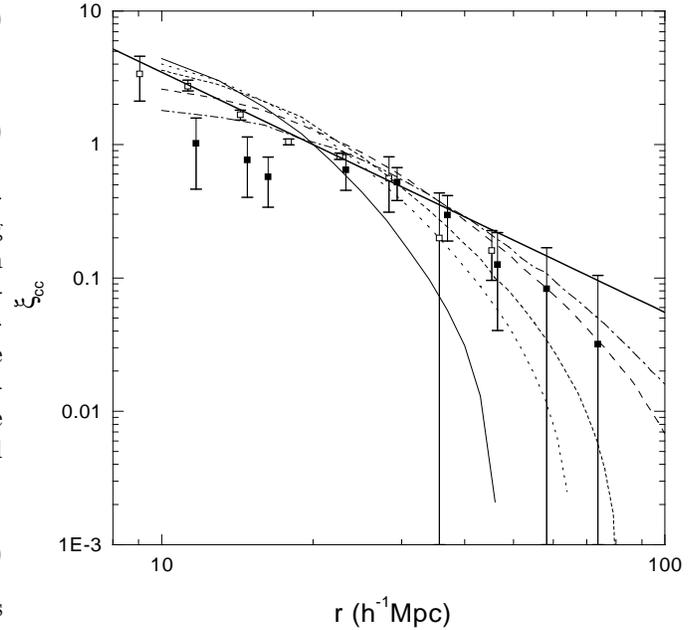,height=9.0cm,width=9.0cm}}
\caption[] {The same as in Fig.8 , but for $\Omega_H/\Omega_M=0.2$}
\label{fig9}
\end{figure}

The cluster-cluster correlation function calculated in this way for
the standard CDM scenario strongly conflicts with the one observed , because it
becomes negative after $\sim 40\;h^{-1}Mpc$. For the standard MDM models with
$\Omega_H\le 0.3$ the situation is better ($\xi_{cc}$ become negative
at $\sim 50\;h^{-1}Mpc$) but positive correlations at $>50\;h^{-1}Mpc$
are not explained by them either (see for example \cite{hol93}; \cite{nov94}).

The positive correlations at $>50\;h^{-1}Mpc$ would be explained in the
cosmological scenario with $\Omega_{M}=1$ and a phenomenological power
spectra with enhanced large scale power (\cite{bar87,ng94,nov96}).
But such spectra are now ruled out by the data
for $\Delta T/T$ at degree angular scales (\cite{SP93})
because their prediction for the rms $\Delta T/T$ is higher than the 95\% c.l.
of the experimental upper limit.

An attractive possibility for avoiding this problem is to consider MLM models
normalised to the COBE quadrupole. The correlation functions of rich clusters of
galaxies calculated as described above agree with
observational data and are positive far beyond $50\;h^{-1}Mpc$ (Figs. 8 \& 9).
As we can see, for MLM models with $0.3\le\Omega_M\le 0.7$,
$\Omega_{H}/\Omega_M=0.1 \simeq 0.2$ and $h=0.5$ the predicted autocorrelation
functions of rich clusters are within the limits of the
observed error bars at large distances. A comparison for other $h$ gives the
following constraints for $\Omega_M h$: $0.13\le\Omega_M h\le 0.35$.

The biasing parameters of rich clusters of galaxies for all models are
in the range $3.3-4.6$ and are in the same range as the values obtained with
different methods from observations (\cite{lyn91,pv91,pl95}).
 The minimal
$\Omega_M h$ is constrained also by the moment of turn around of density
fluctuation peaks which are
associated with rich clusters. Indeed, the requirement that the
total cluster mass collapsed before $z=0$  (not only
the central region but also the frontier areas ) requires
$<\nu>\sigma_{c}\ge 1.06$ (\cite{bar86}), which  is satisfied for models
with $\Omega_M h\ge 0.15$ when $\Omega_{H}/\Omega_M=0.1$ and $\Omega_M h\ge 0.18$ when $\Omega_{H}/\Omega_M=0.2$.

\subsection{Cluster mass function}

The observed abundances of clusters of galaxies is a powerful discriminant
 for different models of dark matter. We will use the  \cite{pss74} (1974, PS)
  formula to compute the number density
  $$
  N(>M,z)= \int_{M}^{\infty} n(M',z) dM',\eqno(17)
 $$
 of virialized objects with
   mass greater than M. According to PS, the comoving number density $n(M,z)$
   of halo masses in the interval $M,M+dM$ , is analytically related to
   the power spectrum by
  $$
  n(M,z)= \sqrt{ \frac {2} {\pi}}
  \frac  {\rho_b(z)}{M^2} \frac {\delta_c} {\sigma(M,z)}\left| \frac {dln\sigma}
  {dlnM} \right |
   \, e^{ -\delta_c^2 / 2 \sigma^2 },\eqno(18)
  $$
here  $\rho_b(z)$ is the background density, $\sigma(M,z)$ is the rms mass
 density fluctuation and $\delta_c$ is the threshold
parameter. The rms $\sigma$ value is defined as
  $$
\sigma^2(M,z)= D^2(z) \int_0^{\infty} \frac {dk}{k} \Delta^2(k) W^2(k,R),
 \eqno(19)
  $$
 where W is a window function and $D$ is the linear growth factor, $D(0)=1$.
The relation between $M$ and $R$ depends on the choice of $W$, for a top-hat
function $R=(3M/4\pi\rho_b)^{1/3}$.

According to linear theory,  $\delta_c=1.686$, for a top-hat window
in an $\Omega_M=1$ universe.
For the more general case $\Omega_M \leq 1 $, $\delta_c$ can be derived
 analytically
 and it has a very weak dependence on $\Omega_M$ (\cite{eke96}).
 On cluster scales, density fluctuations are well described by linear theory
  and so Eq.(18) is thought to be a good approximation to 
  the true number density.
 N-body simulations have been used by various authors
 (\cite{efs88}; \cite{lac94}; \cite{eke96} )
 for an extensive check
  of the validity of the PS predictions. In particular  Eke et al. (1996)
  found that Eq.(18) agrees well with N-body results for CDM models in
  a flat universe.

    For a top-hat choice, the best-fit to N-body results is
     $\delta_c=1.7\pm 0.1$, while for a Gaussian window the $\delta_c$
     threshold is more sensitive to the shape of the power
     spectra (\cite{lac94}).
 In what follows we will take a top-hat window function $W$ and
 $\delta_c=1.686$. Our results will
  be compared with those of Ma (1996) , who has made a similar choice.
  Cluster abundances have been computed for MDM by Ma (1996) ; Bartlett \&
  Silk (1993); Liddle et al. (1996c); Bahcall, Fan \& Cen (1997).
  Clusters of galaxies are rare objects and therefore are very sensitive
  to the value of $\sigma$ .
   Our models have two limiting cases: when $\Omega_{\Lambda} \rightarrow 0$
   ( MDM models) or when $\Omega_{H}/\Omega_M \rightarrow 0$ , which
   corresponds to $\Lambda$CDM models.
For these limiting cases  we have compared our integrations with
published analytical spectra and found
that the greater discrepancies are at high wavenumbers
$ k \magcir 1 h Mpc^{-1}$
and are  $\simeq 10 \% $ of the analytical $P(k)$.

\begin{figure}
{\psfig{file=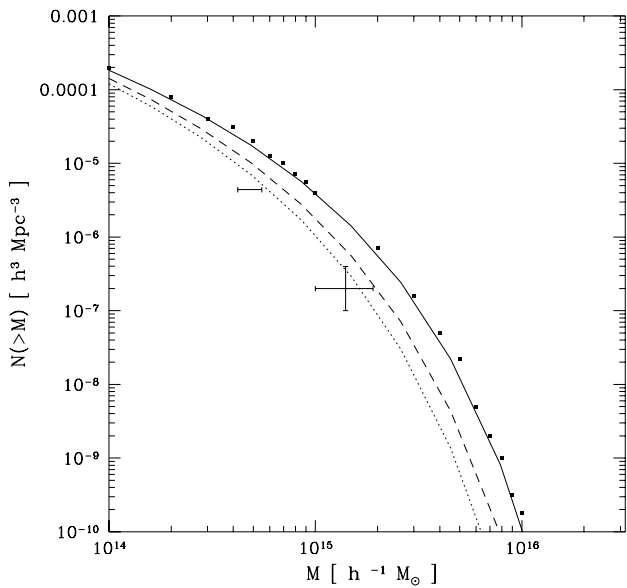,height=9.0cm,width=9.0cm}}
\caption[]{The cluster abundance $N(>M)$ is shown at $z=0$ for MDM
 models with
$\Omega_H/\Omega_M=0.2$, $h=0.5$ and different
values of the $\beta_H$ and $\beta_\nu$ parameters.
The values chosen are:
$\beta_H=2\ \beta_\nu=4$ (solid line),
$\beta_H=6\ \beta_\nu=0$ (dotted short line),
$\beta_H=4\ \beta_\nu=2$ (short-dashed line).
Data points have been taken from Ma (1996, see text). Black squares
 correspond to the integration of Ma for the model
$\beta_H=2\ \beta_\nu=4$.
 The linear spectra used for the calculation of
 $N(>M)$ are normalized according to the COBE 4-yr data}
\label{fig10}
\end{figure}

These differences are mainly due to the absence of baryons in our
 computations. For the length scale of interest to us,
 $ R \simeq 8 -16 h^{-1} Mpc$ and the resulting $\sigma$'s will have similar
 relative errors.
Because of the exponential in Eq.(18), the cluster number density is
strongly affected by
 these errors. We have therefore decided to include baryons in our computations
 of linear spectra. 
 In fact we have obtained  the new spectra
 by taking the computed transfer functions $T(k)$ 
 and interpolating them linearly over a grid of values $q=k/\Omega_M h^2$.

For the same wavenumber $k$ as in the original computation, the new $T(k)$
  is obtained according to the prescription of Sugiyama (1995) :
  $q \rightarrow q/exp(-\Omega_b -\Omega_b/\Omega_M)$.
  We have taken $\Omega_b =0.015/h^2$.
 This procedure works well for CDM transfer functions because, after 
 recombination, baryons
 are caught by the CDM component and their perturbations grow
 together. 
 For spectra with a hot component the ratio $T_H/T_{CDM}$
  to a first approximation does not depend on $\Omega_M$ and the
  prescription can be applied to the total transfer function.
 For MDM Liddle et al. (1996c) have applied a similar procedure to
 the transfer functions of Pogosyan \& Starobinsky (1995), who do not
 include baryons
 in their calculations. They found that the applicability of the
 procedure requires
 $\Omega_b \mincir 0.1$ and $ h \magcir 0.5$, a range of limits that we
 never consider.

 In order to compare our results with observations we take the number density
 of clusters at two mass ranges from the work of Ma (1996).
 These data points correspond to clusters with X-ray temperature greater
 than 3.7 and 7 keV ( \cite{hen91}).
 For the first point White et al.( 1993a) estimated the upper
 limit in mass from the cluster velocity dispersion, while for the lower
 limit  the X-ray temperature of $3.6$ KeV has been converted into a
 mass of $M=4.2\ 10^{14} M_{\odot}$ assuming an isothermal model.
 The second point is taken from
Liddle et al. (1996d) who have used the N-body hydro simulations of
White et al. (1993b) and found $M=(1.23 \pm 0.3) 10^{15} h^{-1} M_{\odot}$
for the virial mass of a cluster with $X$-ray temperature $kT=7$ keV
in a critical universe.
\begin{figure}
{\psfig{file=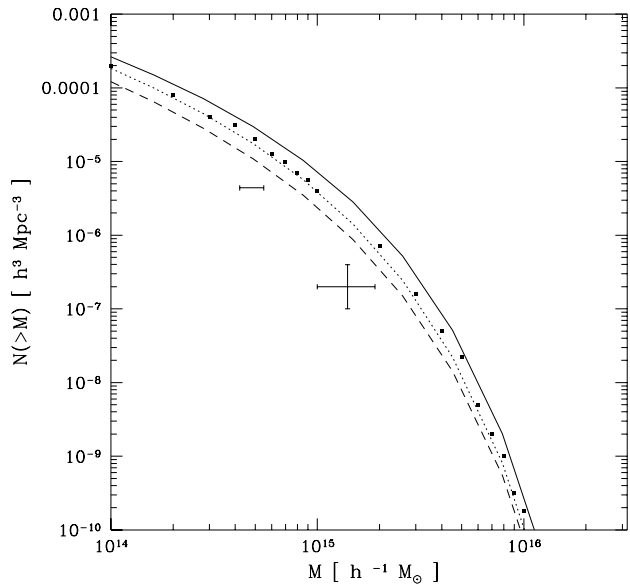,height=9.0cm,width=9.0cm}}
\caption{The same as in Fig. 10 but for MDM models with $h=0.5$,
$\beta_H=2, \beta_\nu=4$, and
$\Omega_H/\Omega_M=0.1$ (solid line),
$\Omega_H/\Omega_M=0.2$ (dotted line),
$\Omega_H/\Omega_M=0.3$ (short dashed line)}
\label{fig11}
\end{figure}

In Fig. 10 we plot the present mass functions of clusters for SMDM
models with $\Omega_H=0.2, h=0.5, \Omega_b=0.06$.
The figure shows the dependence of $N(>M)$ on the number of species of
massless and massive collisionless particles.
 We assume that the total number of species of both particles
is equal to $3$. We can see a weak dependence on these parameters.
The mass function in SMDM models with one species of massive
 neutrinos is above that for models with three massive neutrinos.
 Keeping $\Omega_H$ fixed,  and decreasing the  neutrino mass, then
 $N(>M)$ decreases too.
 It can be seen that the model with $\beta_H=4$, advocated by Primack
 et al. (1995), fares much better.

Fig. 11 shows the mass function $N(>M)$ for SMDM models with
different
values of $\Omega_H=0.1, 0.2, 0.3$ and fixed values of $h=0.5,
\beta_H=2, \beta_\nu=4$.
The black squares correspond to the $\Omega_H=0.2$ case treated
by Ma (1996, figure 8 top right).  With an increase of $\Omega_H$, the
mass function decreases.
We can observe also small differences in the shape of the functions.

As we can see, for the considered range of parameters SMDM models
do not fit cluster abundances, if we adopt the COBE normalization and
spectral index $n=1$ with no gravitational waves.
The $\Omega_H=0.3$ case is marginally consistent, and it was
the one considered by Bartlett \& Silk (1993).
An improvement in the fit can be obtained with the introduction of
a small tilt in the initial spectrum ( $n \simeq 0.9$ ) and/or a tensor
contribution
to COBE anisotropies. This point has been discussed in detail by Ma (1996)
and we do not consider it here.

For MLM models the data points in Fig. 11 should be calculated according to
the model, because the conversion from temperatures to masses depends
 on $\Omega_M$. For this reason we next consider the cumulative cluster
temperature function predicted for different MLM models, and compare it
 with observations.

\subsection{Cluster temperature function}

Identification of clusters of galaxies in the optical band is subject to
 the problems of foreground/background contamination. Projection  effects
 can also undermine mass determination
 through virial analysis (\cite{fre90}; \cite{dck89}).
 On the other hand clusters are also strong X-ray sources (\cite{mck80}),
 their emission does not
  suffer from these problems and clusters can be reliably identified.

  The X-ray emission of galaxy clusters has two physical observable-
  related quantities: the luminosity and the temperature.
During cluster collapse, the gas is shock heated to the virial temperature,
 then it approaches an isothermal distribution in virial equilibrium.
 For the gas temperature $T_g$ one should have $ T_g \propto \alpha M^{2/3}$.
  This relation has been confirmed by numerical simulations
  ( \cite{evr96}; \cite{nav95})
  with a very small dispersion for the coefficient $\alpha$
  and different DM models.

 Then the $T-M$ relation allows us to connect the PS equation (18) to the
 cluster X-ray temperature function (XTF).
We take the Eke, Cole \& Frenk (1996, hereafter ECF) relation for an isothermal gas:
\setcounter{equation}{19}
\begin{eqnarray}
kT_{gas}&=&{7.75 \over \beta} \left({6.8 \over {5X+3}}\right)
{\left({M\over {10^{15}h^{-1}M_\cdot}}\right)}^{2 \over 3} (1+z) \nonumber \\
&&\times{\left({\Omega_M \over \Omega_M(z)}\right)}^{1 \over 3}
{\left({\Delta_c \over 178}\right)}^{1 \over 3} keV, 
\end{eqnarray}

here $\beta$ is the ratio of the galaxy kinetic energy to the gas thermal
energy, $X$ is the hydrogen mass fraction, $\Delta_c$ is the ratio of the mean
halo density within a virial radius to the critical density at the
corresponding redshift. We assume $X=0.76$, $\beta=1$ and $\Delta_c=178$ (ECF).
 For $\Omega_M < 1 $, $\Delta_c$ can be derived analytically
 and is well approximated by $\Delta_c=178 \Omega_M^{0.45}$.

\begin{figure}
{\psfig{file=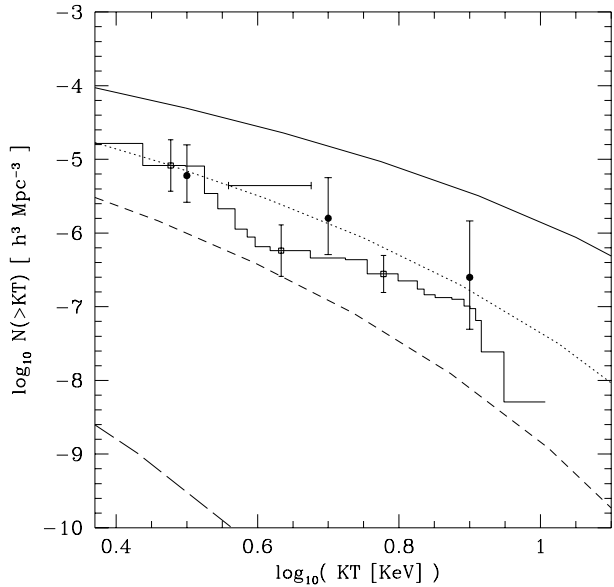,height=9.0cm,width=9.0cm}}
\caption{We show the temperature function of clusters $N(>kT)$
for MLM models with $h=0.5$,
$\beta_H=2, \beta_\nu=4$, $\Omega_H/\Omega_M=0.2$ and different
values of the $\Omega_{\Lambda}$ parameter:
$\Omega_{\Lambda}=0.$ (solid line),
$\Omega_{\Lambda}=0.31$ (dotted line),
$\Omega_{\Lambda}=0.45$ (short dashed line),
$\Omega_{\Lambda}=0.65$ (long dashed line),
$\Omega_{\Lambda}=0.74$ (dot-short dashed line).
The histogram is obtained from the X-ray data set of  Henry and Arnaud (1991).
Open squares are three selected temperatures for which error bars have been
estimated with a bootstrap procedure. Black circles refer to the mass function
of Bahcall and Cen (1993). The horizontal bar is from White et al.(1993a) }
\label{fig12}
\end{figure}
 For a specified model Eqs.(18-20) allow us to compute the cluster
 XTF $N(> kT)$. Then this function can be compared with present data
 and can be used to constrain different DM models (\cite{bar93}; ECF;
 \cite{via96}).
 Current observations from Einstein and EXOSAT satellites have been
 used to determine the local cluster XTF (\cite{hen91}; \cite{edg90}).
 Our study of the cluster XTF will make use of the Henry \& Arnaud
 (1991) data. In particular we will closely follow the analysis of
 ECF, this will make it easy to compare of our MLM predictions
 with previous result for $\Lambda$CDM models (ECF).

 The estimated cumulative cluster XTF is computed according to
 $$
 N(> kT)=\sum_{T_i>T} 1 / V_{max,i}\ , \eqno(21)
$$
 here $V_{max,i}$ is the maximum volume at which the $i-th$ cluster can be
 detected for a specified flux limit
 ($F_X=3\,10^{-11} erg sec^{-1} cm^{-2} $) in
 the $2-10 $ Kev range.

The cumulative XTF obtained in this way is shown as the solid line histogram
 in Fig. 12.
Error bars
 have been found with a bootstrap procedure, applied to the original
 sample of 25 clusters, at three different temperature bins $T_j$.
The resulting $1 \sigma$ amplitudes $\delta_j $ are plotted as open squares
in Fig. 12, at the three corresponding temperatures.
The $\delta^2_j$ are defined as $\delta^2_j =
< (\log_{10}N^{boot}(>kT_j) -\overline{\log_{10}N^{boot}(>kT_j)})^2>$, the average being over the bootstrap ensemble.

 The estimated $N(> kT)$  agrees with the one obtained by ECF
 (see their Fig. 3),
 the procedure being the same, and is consistent with the differential one
 of Henry \& Arnaud (1991), although ECF pointed out some errors
 in the original estimate that almost cancel each other.

  The horizontal bar on the left upper part is the first data point
  of the previous $N(>M)$ figures.
  The point was obtained from White et al.(1993a) precisely from the
  temperature function of Henry \& Arnaud (1991) and it
  proves the consistency of our procedure.
 The small offset along the $kT$ axis can be attributed to the
 use of the differential or cumulative temperature functions (ECF).

 A different  test is to compare our $N(>kT)$ with the cluster
 mass function of Bahcall \& Cen (1993). Their best fit $N(>M)$ is plotted
 as black circles for three different temperatures.
 In order to convert masses into temperatures we have used Eq.(5) of
 Bahcall \& Cen (1993). Error bars represent the $ 1\sigma $ dispersion
 of their Eq.(5). As can be seen, there is a substantial agreement between
 the two estimates.

In comparing the predictions of our models with the observed XTF one
must be aware of  possible uncertainties in the mass-temperature
relation. Because of the steep decline of the XTF with temperature,
even a small dispersion can have drastic effects.
In Eq.(20) the parameter $\beta$ represents the ratio between the
virial and gas temperatures (we have assumed an isotropic isothermal
profile). The $\beta$ parameter can be calculated either from the data
or through numerical simulations.

Current available data are consistent with $\beta \simeq 1$
(\cite{edg91}; \cite{squ96}).
The hydrodynamical simulations of Navarro, Frenk \& White (1995)
  give a result of $\beta \simeq 1$ and
 have been used by ECF , who assume $\beta=1 \pm 0.1$.
This scatter around unity for $\beta$
( $\simeq 10 \%$) has been confirmed in more
recent work by Eke, Navarro \& Frenk (1997, see also \cite{evr96}), who have
performed a set of N-body hydrodynamical simulations
to investigate the X-ray evolution for a set of clusters in a
low-density flat CDM cosmology.

Thus we have reliably assumed $\beta=1$ in order to convert masses
into temperatures.
The theoretical XTF has then been computed from Eq.(18), for MLM
models with different $\Omega_{\Lambda}$  and a fixed ratio
$\Omega_H /\Omega_{M}=0.2$. We normalize the final spectra
according to the COBE data (see sect.2.2 ). The results are shown in Fig. 12.
\begin{figure}
{\psfig{file=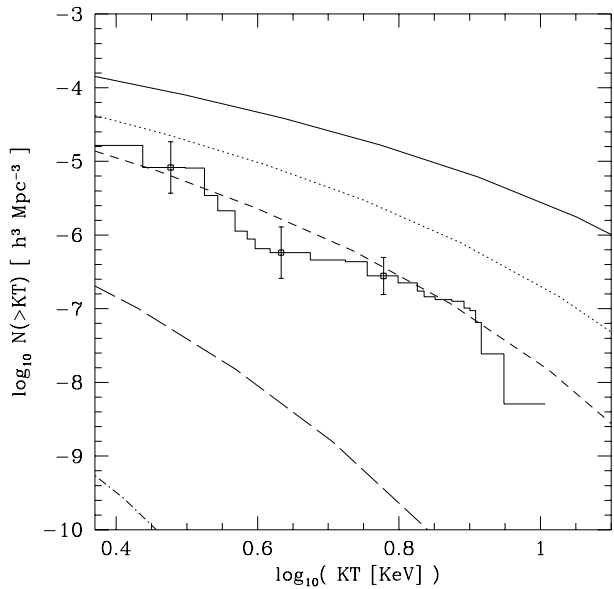,height=9.0cm,width=9.0cm}}
\caption{The same as in Fig.12, but for $\Omega_H/\Omega_M=0.1$}
\label{fig13}
\end{figure}

The inclusion of a $\Lambda $ term clearly alleviates the problem for
MDM and brings the models in good agreement with the data.
The solid line is the limiting case $\Omega_{\Lambda}=0$, which
corresponds to the standard MDM. In order to fit the present cluster
abundance, this model would
require a value $\beta \simeq 2$, clearly inconsistent with
present estimates.

\begin{figure*}[!ht]
\centerline{\hbox{%
\psfig{figure=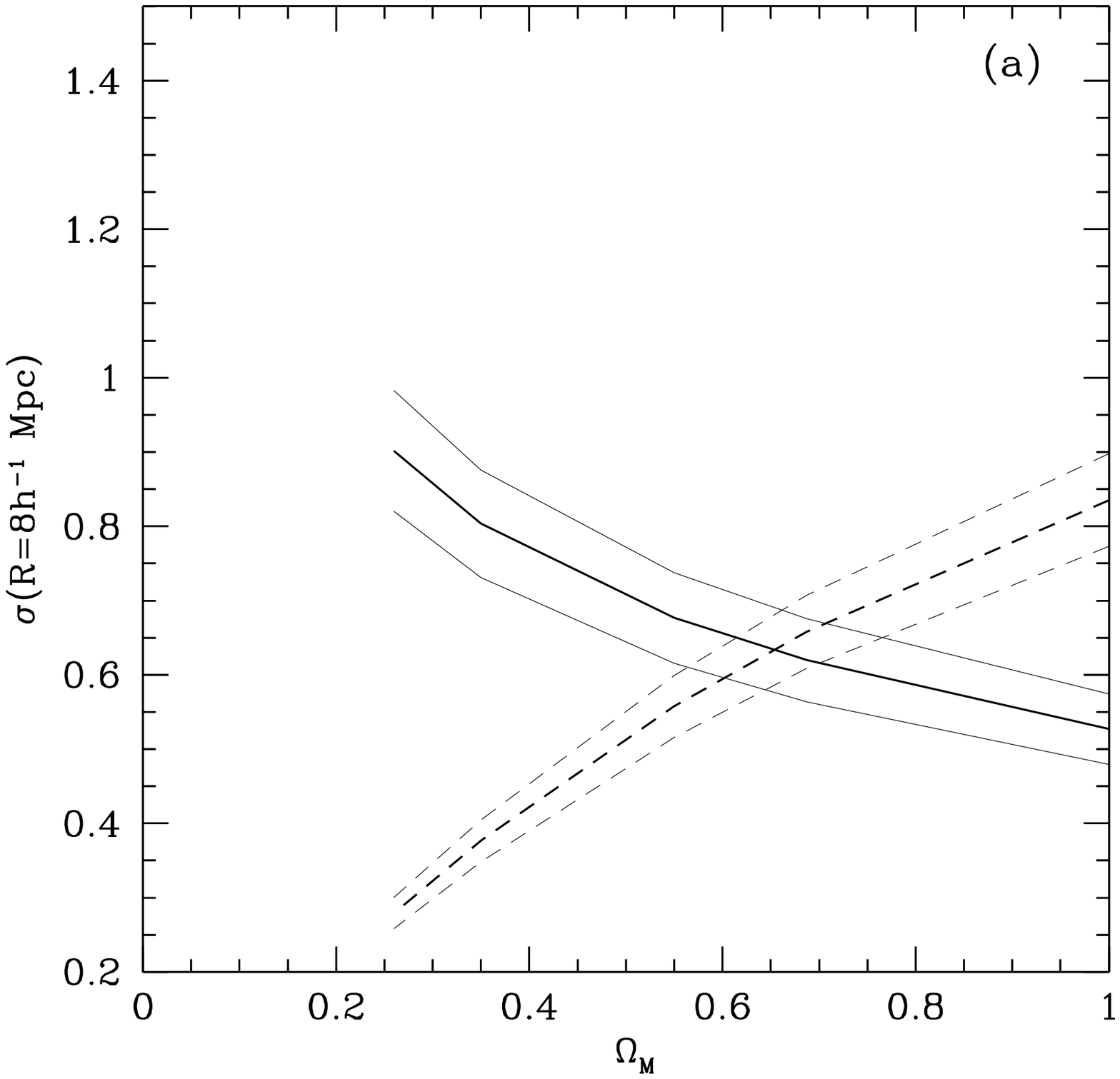,height=8truecm,width=8truecm}%
\psfig{figure=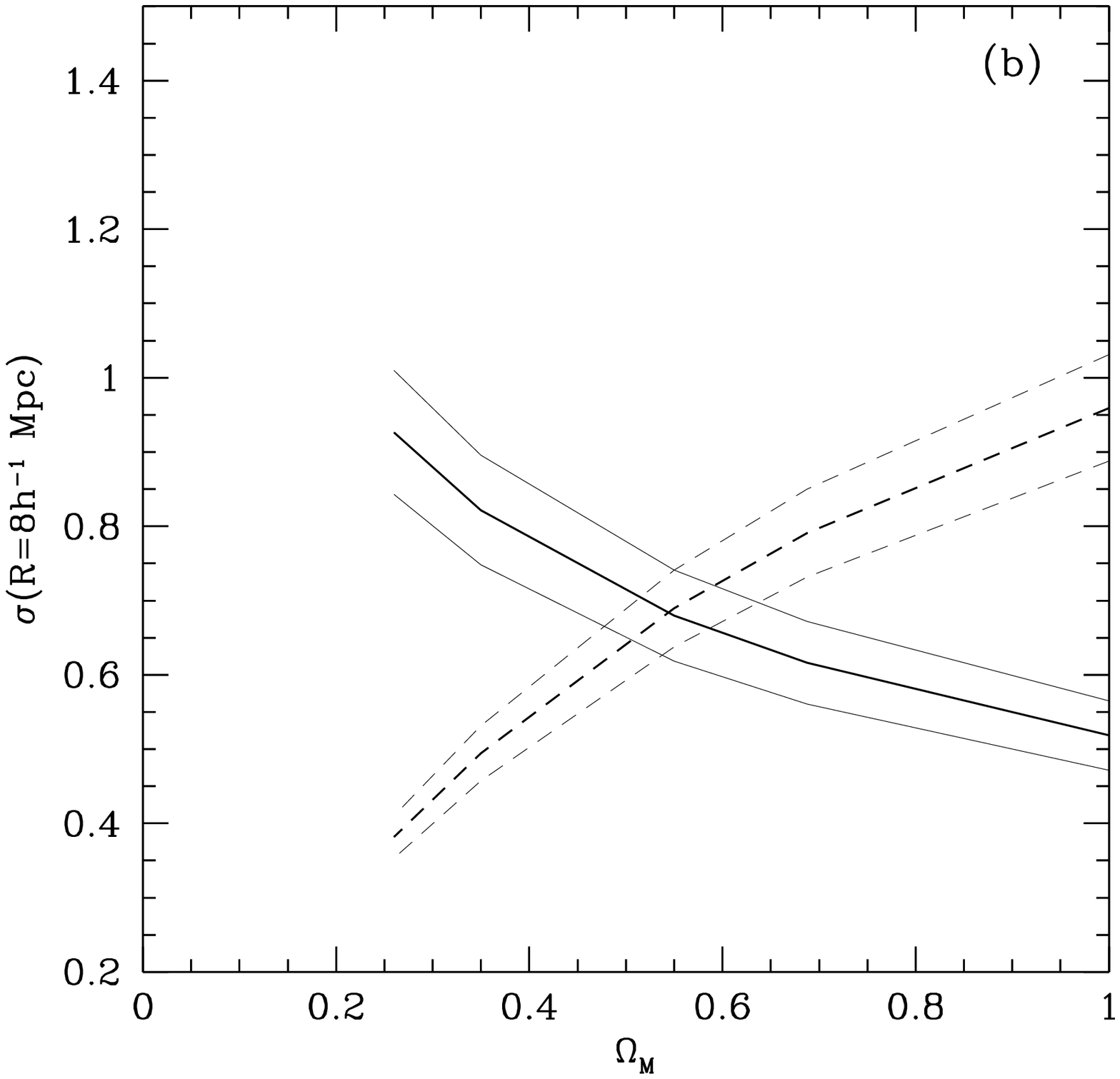,height=8truecm,width=8truecm}%
}}
\caption[]
{{\bf a} $\sigma_8$ is shown as a function of $\Omega_M$ for MLM models 
with $h=0.5, \beta_H=2, \beta_\nu=4$, $\Omega_H/\Omega_M=0.2$.
The continuous line is the best-fit to the estimated cluster  abundance,
the dashed line is from the COBE 4-yr data. The thin lines show the statistical
uncertainties, quoted in the text. {\bf b} The same as in panel {\bf a}, but 
for $\Omega_H/\Omega_M=0.1$}
\label{fig14}
\end{figure*}

From Fig. 12 the best range of $\Omega_{\Lambda}$ for fitting the data
is $\Omega_{\Lambda} \simeq 0.3-0.5$.
Fig. 13 shows the same plot of Fig. 12 , but for
$\Omega_H / \Omega_M =0.1$.
  In this case the best range for
$\Omega_{\Lambda}$ is $ \simeq 0.4 -0.6$.
 In order to constrain a particular
 model we can now compute the $\chi^2$ quantity
$$
\chi^2= \sum_{j=1}^3 \sum_{k=1}^3 y_j C_{jk}^{-1} y_k , \eqno(22)
$$
where $j$ is the bin index, $y_j=(\log_{10}(N_j)-\log_{10}(N_j^{th})) $,
$N_j=N(>k T_j)$ and $N_j^{th}$ is for a
particular model from Eq.(18).
The covariance matrix $C_{jk}= < \delta_j \delta_k>$ takes into account
the correlations between different temperature bins.
According to ECF these correlations are
not negligible, but the models which minimize $\chi^2$  do not depend
strongly on their inclusion in Eq.(22).
Thus ECF take for $C_{kl}$ its diagonal form.
 However the models that we consider have different spectra and we have
chosen to keep the whole matrix for the minimization of Eq.(22).

From Eqs.(18-19) $\chi^2$ is a function of the power spectrum
constant $A$ or, equivalently, of the rms mass fluctuation
$\sigma_8 \equiv \sigma(R=8h^{-1})$.
From the minimization of Eq.(22) the formal error on $\sigma_8$ is 5 \%, but
uncertainties on the other parameters will affect $\sigma_8$ too.
From N-body integrations ECF assume a scatter of 4 \% for $\delta_c$.
This is not surprising because for the length scale of interest to us we
can neglect tidal forces and assume that the collapse is spherically
symmetric. Other sources of errors are the scatter in $\beta$ ( \mincir 10 \% )
, the sample completeness ( 90 \% at $F_X=3\ 10^{-11} erg sec^{-1} cm^{-2}$,
\cite{lah89} ) and errors in the measurement of temperatures.
 For the latter error, ECF analysis gives an upper limit of  \mincir 1 \%.
 Summing all these errors in quadrature
 the final dispersion for $\sigma_8$ is about 10 \%,
 twice the statistical error.

 For a given model we can now estimate, from Eq.(22), the $\sigma_8$ which
 is consistent with the estimated cluster abundance, and compare it
 with the $\sigma_8$ obtained from COBE data. These two values for $\sigma_8$ 
 will
 in general not coincide, however there will be a set of values of cosmological
 parameters for which they will be in the same range.
  The estimated uncertainties allow us to judge the reliability
  of the overlap.
For the COBE normalization we have assumed a 7 \% statistical error
(\cite{bun97}).

We show  in Fig. 14a the $\sigma_8$ ( continuous line ) which is obtained
from the minimization of the  $\chi^2$ quantity. The dashed line is
 from COBE. The two sigmas  are plotted as function of the cosmological
 parameter $\Omega_M$.
The set of models is for $h=0.5$ and $\Omega_H/\Omega_M=0.2$.
 Thin lines represent the assumed uncertainties.

From Fig. 14a the best range of $\Omega_{M}$ which fits the data
is $0.55 \mincir \Omega_{M} \mincir 0.75$.
 The standard MDM model ($\Omega_{M}=1$) is rejected at the $2\sigma$
  level.
Fig. 14b shows the same plot as in panel (a), but for
$\Omega_H / \Omega_M =0.1$. In this case the best range for
$\Omega_{M}$ is $0.45 \mincir \Omega_{M} \mincir 0.65$.
 In Fig.15 we consider the same set of models but for $h=0.7$.
 We obtain $0.3 \mincir \Omega_{M} \mincir 0.5$
 for $\Omega_H / \Omega_M =0.2$ and $0.3 \mincir \Omega_{M} \mincir 0.4$
 when $\Omega_H / \Omega_M =0.1$.

\begin{figure*}[!ht]
\centerline{\hbox{%
\psfig{figure=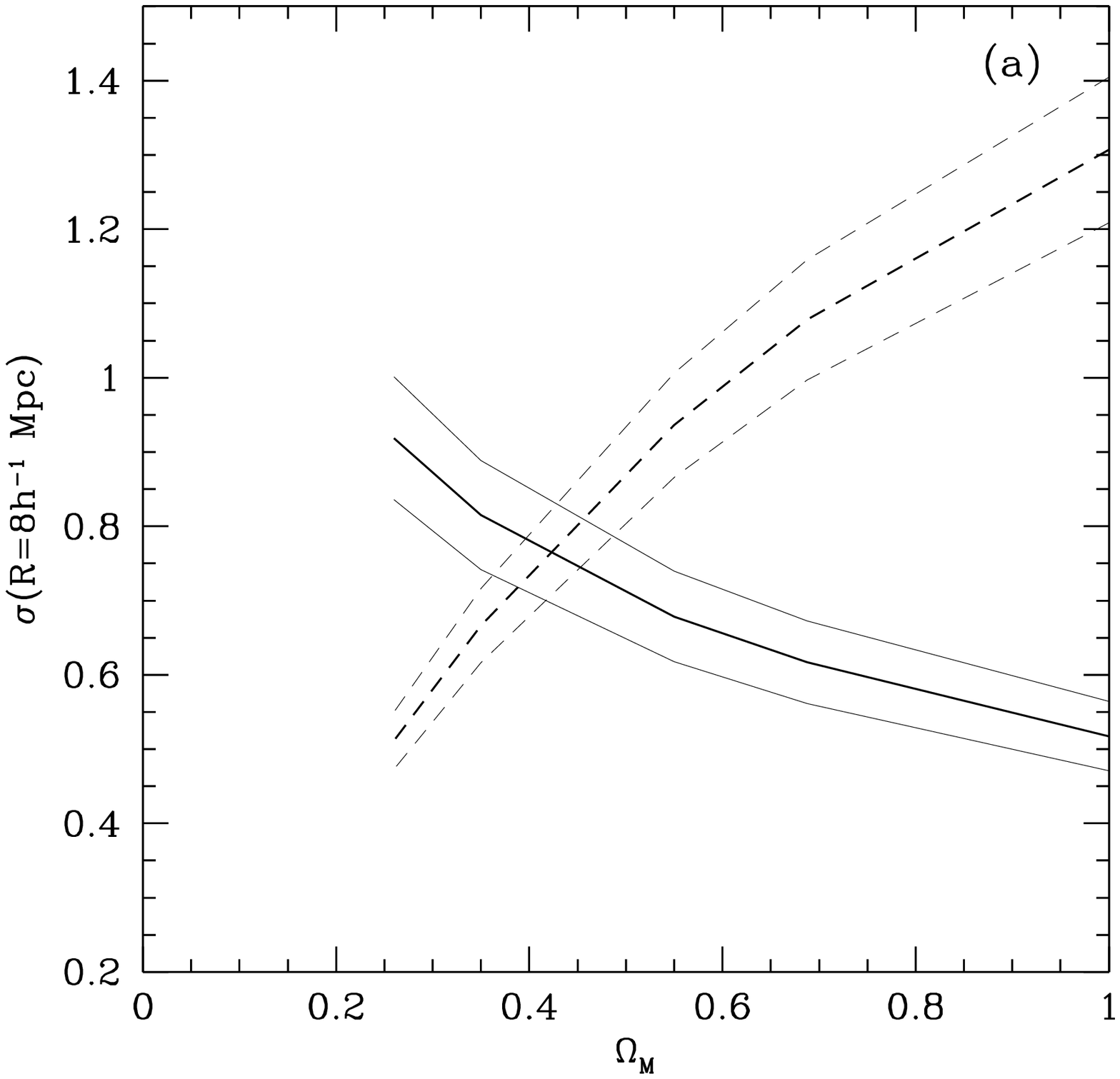,height=8truecm,width=8truecm}%
\psfig{figure=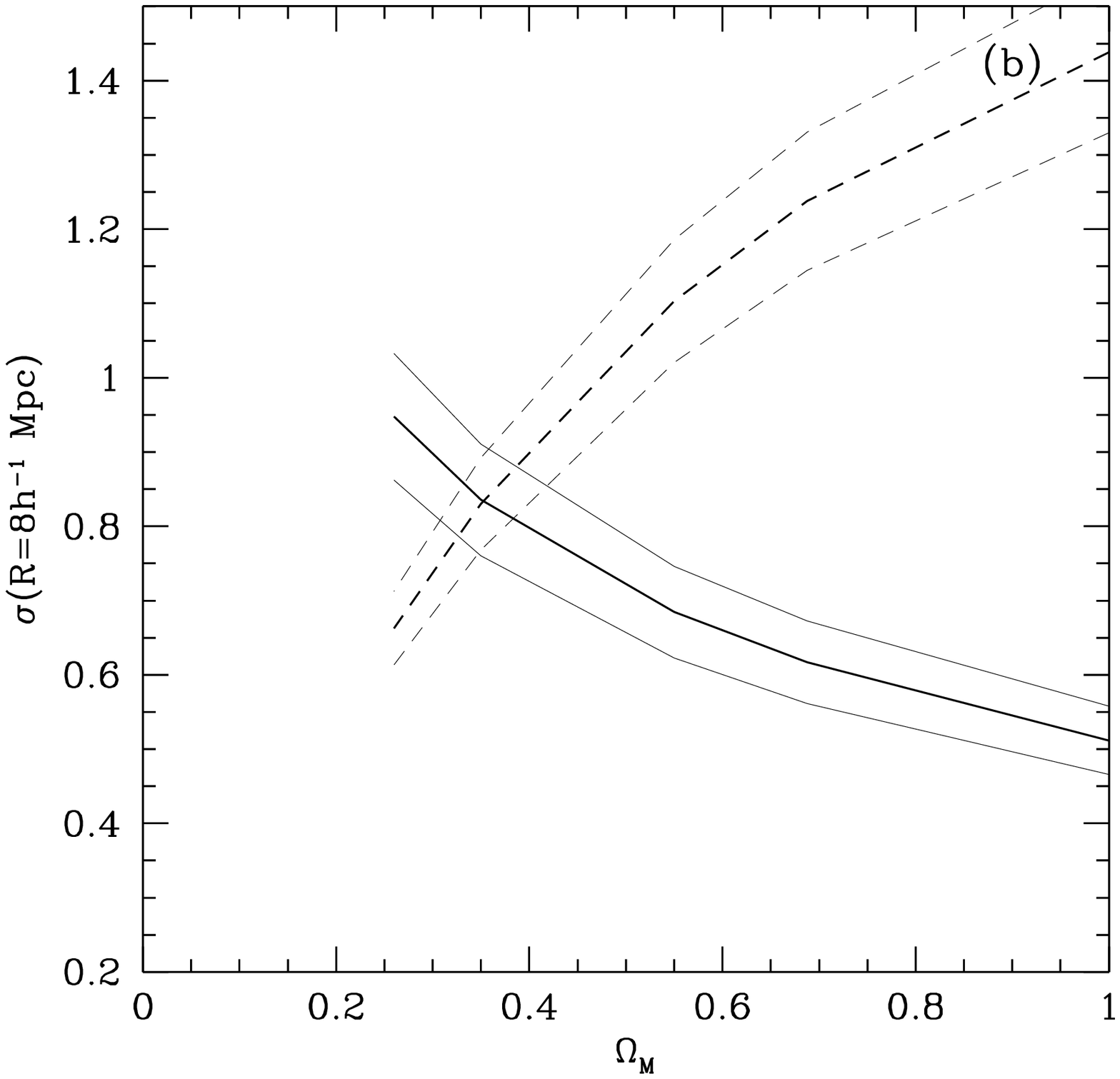,height=8truecm,width=8truecm}%
}}
\caption[]
{{\bf a} $\sigma_8$ is plotted versus $\Omega_M$ for MLM models 
  with $h=0.7, \beta_H=2, \beta_\nu=4$, $\Omega_H/\Omega_M=0.2$.
 The meaning of the lines is the same as in Fig.14. {\bf b}
 The same as in panel {\bf a}, but for $\Omega_H/\Omega_M=0.1$}
\label{fig15}
\end{figure*}

 According to the value of $h$ we can then summarize the following constraints
  for MLM models  with one species of massive neutrinos and
  $\Omega_H / \Omega_M \le 0.2$:
$ 0.5 \mincir \Omega_{M} \mincir 0.8$ ($h=0.5$) or
$ 0.3 \mincir \Omega_{M} \mincir 0.5$ ($h=0.7$).
 The corresponding allowed values of $\sigma_8$ are :
 $\sigma_8 \simeq 0.65 \pm 0.1$  ($h=0.5$) or
 $\sigma_8\simeq 0.8 \pm 0.1$ ($h=0.7$).
 A comparison of these constraints with the clustering data of
 sect.4.1 must take into account the biasing factor. A rescaling of 
 $\Delta^2(k)$ according to $ b_{gal}=1/\sigma_8$ shows that
consistency with cluster abundance  is marginal for models
 with $\Omega_H / \Omega_M = 0.2$. For $h=0.5$ and
$\Omega_H / \Omega_M = 0.1$ we obtain 
 $ 0.55 \mincir \Omega_{M} \mincir 0.65$ , while 
 for $h=0.7$ the limits on $\Omega_M$ are  
 $ 0.35 \mincir \Omega_{M} \mincir 0.4$.

The constraints on $\Omega_M$ that we obtain  
follow from comparing our results with the present cluster XTF. 
We have not considered possible constraints that might  be given 
by considering cluster evolution.
 The evolution of cluster number density with redshift is a powerful
 tool for discriminating among different cosmologies.
This follows directly
 from the different growth of density fluctuations  in different
 cosmologies.  The fluctuations in models with $\Omega_M < 1$ grow
very slowly and structures will experience little evolution at
 recent times. On the contrary, for $\Omega_M=1$, the higher growth
 rate of density fluctuations implies for galaxy clusters a much stronger
 evolution  at late redshifts.

 Bahcall, Fan \& Cen (1997) compared the results for cluster
 abundance, using large-scale  N-body simulations, with recent data 
 at $z\simeq 0.5-1$. They found that MDM models are ruled out at the $2\sigma$
 level, while a $\Lambda$CDM model with $\Omega_M=0.34\pm0.13$ $h=0.65$ can
 consistently fit the data. These simulations show that the cluster evolution
 rate is  strongly model dependent. We accordingly do not attempt here to 
 perform 
 a linear analysis, applying Press-Schechter theory, to compute the redshift
 evolution of the cluster number density for MLM models.
  
 We suggest then that large scale numerical simulations can be used
 to compare the cluster evolution for MLM models with recent data.
 These tests are likely to strengthen or falsify the models, given the
 already narrow window of allowed values for the cosmological parameters
 arising from the linear tests applied here.

\subsection{Damped Lyman-$\alpha$ systems}
 An important test for dark matter models with a massive neutrino
 component is given from the observations  of objects at high redshifts.
 Due to free-streaming effects the dark matter power spectrum will
 be severely damped on small scales, thus making formation of early
 objects more difficult than in a model without the hot component.
 The most important class of such objects are damped Lyman-$\alpha$ absorption
 systems.

 These objects have a high column density of neutral hydrogen ($N_{HI} \magcir
 10^{20} cm^{-2}$) and are detected by means of absorption lines in quasar 
 spectra (Wolfe 1993).
Observations at high redshift have lead to estimates of the abundance of neutral
 hydrogen in damped Lyman-$\alpha$ systems 
 (Lanzetta, Wolfe \& Turnshek 1995; Storrie-Lombardi et al. 1996).
The latter authors have analyzed spectra at $z=3$ and $z=4$ , we will make use
of the $z=4$ data, which gives the strongest constraint on the
primordial spectrum.
According to Storrie-Lombardi et al. (1996), the cosmological mass
density of neutral hydrogen gas $\Omega_g$ is given by
$$
\Omega_g(z=4)=(0.0011 \pm 0.0002)h^{-1} \sqrt{ \frac {\Omega_M} {\Omega_M(z=4)}},\eqno(23)
$$
with the original data being given in units of the critical density, the
square root taking into account different cosmologies.

The standard view is that damped Lyman-$\alpha$ systems are a population of 
protogalactic disks (\cite{wof93}), with a minimum mass of
 $ M \ge 10^{10} h^{-1} M_{\odot}$ (\cite{hae95}).
 The fractional density of collapsed objects of minimum mass $M_{DLAS}$ is
  then 
\setcounter{equation}{23}
\begin{eqnarray}
 \Omega_{DLAS}(z=4)&=&\frac {\Omega_g(z=4)} { f_{gas} \Omega_b}=
(0.069 \pm 0.021) \frac {h} {f_{gas}} \nonumber \\
&&\times \sqrt { \frac {\Omega_M} {\Omega_M(z=4)}}
\end{eqnarray}

where $f_{gas}$ is the fraction of neutral hydrogen and $\Omega_b = 0.016h^{-2}$
 (\cite{cop95}). In Eq. (24) the error in $\Omega_b$ has been added in
 quadrature (Liddle et al. 1996c).
 A conservative assumption is $f_{gas}=1$, but recent hydrodynamical simulations
  (\cite{mae97}) have claimed $f_{gas}\le 0.1$.
  We will consider constraints on our models arising from both of the limits on 
  $f_{gas}$.
\begin{figure}[!ht]
{\psfig{file=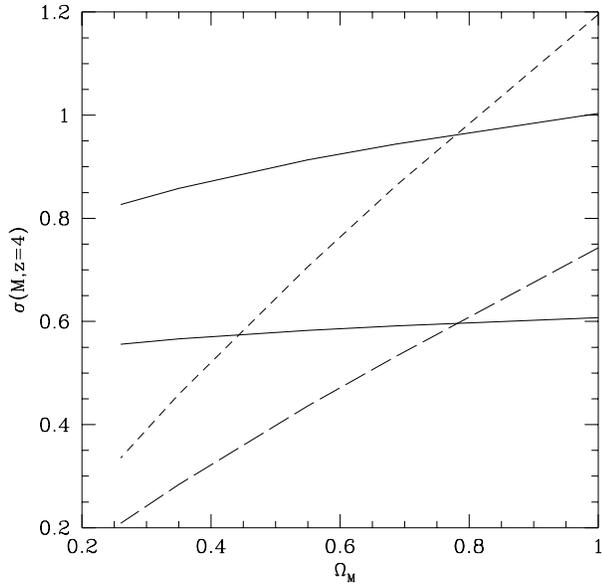,width=9.0cm}}
\caption[]
{$\sigma(M,z=4)$ is shown as a function of $\Omega_M$ for $h=0.5$.
The continuous line represents the 95 confidence level for the 
 the lower limit obtained from observational data using Press-Schechter
 theory. The bottom line is for $f_{gas}=1$ and the top line refers to
 $f_{gas}=0.1$. The $\sigma(M,z=4)$ for MLM models are plotted as dashed
  lines for different values of the ratio $\Omega_H/\Omega_M$.
 The long(short)-dashed line is for $\Omega_H/\Omega_M=0.2 (0.1)$}
\label{fig16}
\end{figure}

 The theoretical counterpart of Eq.(24) can be found using the Press-Schechter
 (1974) theory and is given by
 $$
 \Omega_{DLAS}( > M, z) = {\rm erfc} \left [ \frac {\delta_c}{ \sqrt {2} \sigma(M,z)}
 \right ], \eqno(25)
 $$
 where erfc is the complementary error function and $\sigma(M,z)$ is defined 
 according to Eq.(19) using a top-hat window function.
 Particular care must be taken when inserting the minimum mass into Eq.(19)
 because for damped Lyman-$\alpha$ systems $M=M_{DLAS}$ is well below the
 neutrino clustering scale at $z=4$.
 We correct $M_{DLAS}$ according to $ M_{DLAS} \rightarrow 10^{10} h^{-1}
 (1 -\Omega_H/\Omega_M)^{-1} M_{\odot}$ (Liddle at al. 1996c).
 The collapsed state of damped Lyman-$\alpha$ systems is uncertain and a 
 conservative assumption is that these systems have collapsed along two
 axes. We accordingly take for the threshold parameter  the value
 $\delta_c=1.5$ (Monaco 1995).

 We conservatively assume the theoretical predictions (25) to be bounded
 from below by Eq.(24), for $\Omega_{DLAS}$ we take the $2\sigma$ lower
 limit. The rms mass fluctuation $\sigma(M,z=4)$ can now be constrained
 from below using Eqs.(24-25).
 We show these limits as a function of $\Omega_M$ in Figs. 16 \& 17, for
 $h=0.5$ and $h=0.7$, respectively.
 In these figures the bottom solid line is for $f_{gas}=1$, while the top
 solid line refers to $f_{gas}=0.1$. This value gives a much more severe 
 constraint
  on $\sigma(M,z=4)$ than for $f_{gas}=1$.
  The theoretical values  of $\sigma(M,z=4)$ are shown for 
  $\Omega_H/\Omega_M=0.1$ ( short dashed line ) and
  $\Omega_H/\Omega_M=0.2$ ( long dashed line ).
At a given $\Omega_M$ the value of $\sigma(M,z=4)$ decreases as the ratio
  $\Omega_H/\Omega_M$ increases. This behavior of $\sigma(M,z=4)$ follows
  because an increase of $\Omega_H$ implies, for a given
  normalization and redshift, a reduction in the available power at
  small scales.
 The effect becomes more pronounced for $\Omega_M \rightarrow 1$ 

 From Fig. 16 one can infer that for $h=0.5$ MLM models 
  with $\Omega_H/\Omega_M = 0.1$ 
are not strongly constrained by the available data for 
damped Lyman-$\alpha$ systems.
  For $h=0.5$ and $\Omega_H/\Omega_M = 0.2 (0.1)$ we obtain
the following lower limits :  $ 0.8 (0.4) \mincir \Omega_M$.
For $\Omega_H/\Omega_M = 0.2 $ this lower limit on $\Omega_M$ do not overlap 
with the upper limit given by cluster abundances and the model is 
inconsistent.
 For $h=0.7$ we obtain from Fig. 17~$0.4(0.2) \mincir \Omega_M$ if
   $\Omega_H/\Omega_M = 0.2 (0.1)$.
 This lower limit is consistent with the upper limits
 from cluster abundances and other constraints for the same set of models.

These constraints on $\Omega_M$ have been obtained for a minimum mass
 of $ M = 10^{10} h^{-1} M_{\odot}$, the role of possible uncertainties
 on the limits for $\Omega_M $ can be considered by decreasing this
 lower limit by an order of magnitude.
 For $ M = 10^{9} h^{-1} M_{\odot}$ we obtain new constraints on $\Omega_M$ 
which do not differ in a relevant way from those shown in Figs. 16 \& 17.
\begin{figure}[!ht]
{\psfig{file=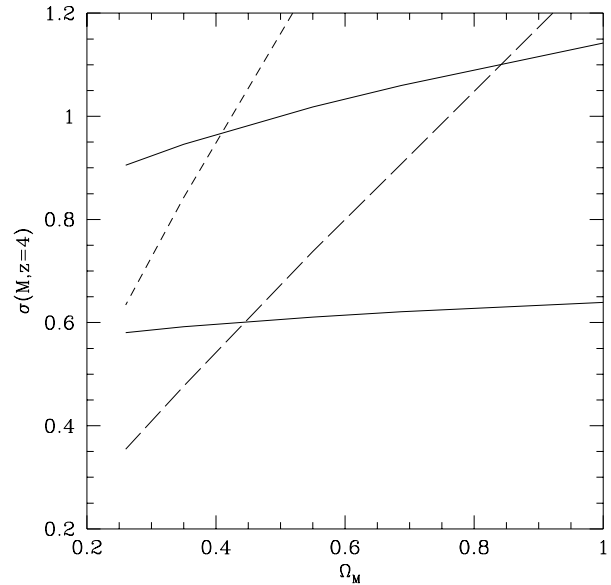,width=9.0cm}}
\caption[]
{ The same as in Fig.16, but for $h=0.7$} 
\label{fig17}
\end{figure}

 We have considered $f_{gas}=1$ but if we take $f_{gas}=0.1$ then
  none of the MLM models for $h=0.5$ survives the constraints from damped 
  Lyman-$\alpha$ systems.
 The model with $\Omega_H/\Omega_M = 0.2 $ is excluded 
and $\Omega_H/\Omega_M = 0.1$ requires $\Omega_M\magcir 0.8$, a minimum value
excluded at the $1\sigma$ level by cluster abundances.
 Also MLM models with $h=0.7$ are severely constrained if $f_{gas}=0.1$.
In this case we have $\Omega_M \magcir 0.8 (0.4)$ for 
   $\Omega_H/\Omega_M = 0.2 (0.1)$.
 Thus the model with 
   $\Omega_H/\Omega_M = 0.2 $ is totally inconsistent with cluster abundances
   but $\Omega_H/\Omega_M = 0.1$ is still within the $1\sigma$ 
   range.
 We conservatively take the lower limit on $\Omega_M$ arising from 
 $f_{gas}=1$. Numerical hydrodynamical simulations have tested only MDM models
 (\cite{mae97}), hydro simulations with MLM spectra are clearly required
in order to  obtain a tight limit on $f_{gas}$.

\section{ Conclusions}

We have discussed linear clustering evolution  for a set of spatially
flat MDM models with a cosmological constant.
We have not considered the role of gravitational waves or of
a possible tilt in the primordial spectrum.
In order to restrict the range of allowed values for the cosmological
parameters, we have applied linear perturbation theory
to compare the predictions of our MLM models with a set of linear data.
The models considered had one species of massive neutrinos
and a scale-invariant spectrum.

 Linear calculations for the dimensionless power spectra
 $\Delta^2(k)$ have been compared with the reconstructed
 real-space power spectrum of Peacock \& Dodds (1994), Peacock (1997).
 Because we have considered linear spectra evolution
 a comparison of clustering is meaningful only at low $k$
  ($ \mincir 0.2h Mpc^{-1})$
A substantial agreement is obtained for those 
unbiased 
models with
$\Omega_H/\Omega_M \mincir 0.2$, and $\Omega_M \magcir 0.6$
($ h = 0.5 $) or
  $ 0.3 \mincir \Omega_M \mincir 0.5$ ($h=0.7$).

The computed linear power spectra have been filtered with a Gaussian
 window of radius $R_f=12h^{-1}Mpc$ to calculate the peculiar velocity
 field. A comparison of the rms bulk motion with POTENT data
 does not lead to strong constraints : $\oml \mincir 0.7$ and
 $\Omega_H /\Omega_M \mincir 0.2$.
 The cluster-cluster correlation functions calculated for
 MLM spectra in the framework of Gaussian random density fluctuation
 field explain the observed positive correlations at $>50\;h^{-1}Mpc$
 and are within the limits of the error bars for that observed at
 large distances when $0.13\le\Omega_M h\le 0.35$.

 Of the considered tests the most important turns out to be that based
 on cluster abundances, for which the present X-ray cluster temperature function
 can put strong limits on the initial spectrum amplitude, or conversely
 on $\sigma_8$.
 Because of the strong dependence of $N_{cl}(> kT)$ on temperature,
  the conversion of X-ray temperatures into
  masses is critical . From the discussion of Section 4.5 this conversion can be
  obtained with a dispersion of less than 10 \%. We have then used
 the Press \& Schechter (1974) formalism to compute the cluster number density
 for different MLM models.
 Consistency with present data  can be achieved for COBE normalized
  models only for the following range of values 
  (at $1\sigma$ level): $\Omega_H /\Omega_M \le 0.2$,
 $ h =0.5(0.7)$ and $ 0.45(0.3) \le \Omega_M \le 0.75(0.5)$.
 In correspondence with these limits, the allowed values of 
 $\sigma_8$ are :
 $\sigma_8 \simeq 0.65 \pm 0.1$  ($h=0.5$) or
 $\sigma_8\simeq 0.8 \pm 0.1$ ($h=0.7$).
 These limits suggest that MLM models require a moderate amount 
 of bias to be consistent with clustering data.
For these range of values consistency with clustering data 
yields $ 0.55(0.35) \le \Omega_M \le 0.65(0.4)$
for $ h =0.5(0.7)$ and $\Omega_H /\Omega_M = 0.1$.
Models with $\Omega_H /\Omega_M = 0.2$ are almost inconsistent.

  One can also compare these constraints with those given 
  by considering cluster evolution.
 The evolution of galaxy clusters has recently been used also for determining
 the cosmological parameters $\sigma_8$ and $\Omega_M$.
 Fan, Bahcall \& Cen (1997) found $\sigma_8=0.83\pm 0.15$. This range of values
 is consistent with the $\sigma_8$ required for MLM with $h=0.7$
 (Fig. 15), and marginally consistent if $h=0.5$.
 At the $2\sigma$ level none of the MLM models considered here is
 ruled out.
 Finally Henry (1997) has used recent data for the evolution of 
the  cluster X-ray temperature function to 
 determine $\Omega_M$ and $\sigma_8$. For a flat cosmology he obtains 
$\Omega_M=0.55 \pm 0.17$ and $\sigma_8=0.66^{+0.34}_{-0.17}$, in
close agreement with our findings for MLM models.

The observed abundance of damped Lyman-$\alpha$ systems has been
 used to obtain lower limits on the rms mass fluctuation $\sigma(M)$ at
  $z=4$ using Press-Schechter theory.
These limits yield the following constraints on $\Omega_M$ 
for MLM models with $\Omega_H /\Omega_M \le 0.2$ :
$0.8-0.4 (0.3) \mincir \Omega_M $ for $h=0.5$ ($h=0.7)$.
 These values have been obtained assuming the fraction of neutral 
 hydrogen to be unity. If this fraction is close to $10 \%$, as suggested by
 numerical hydro simulations, then the constraints for 
 MLM models become much more severe.
 In this case consistency with both damped Lyman-$\alpha$ and cluster abundances
 can be achieved only for $h=0.7$ , 
$\Omega_M \simeq 0.4$ and $\Omega_H /\Omega_M =0.1$.

In order to further restrict our models we can also consider present
observational constraints on cosmological parameters.
 The cosmological constant is constrained to be $\oml \le 0.66$ at
  the 95 \% confidence limit by QSO lensing  (Kochanek 1996).
  A tighter restriction comes  from the recent work of Perlmutter et al.
  (1997) on SN Ia , who give $\Omega_{\Lambda} < 0.51$ still at 95 \%
 c.l. .

 Recent Hipparcos data ( Feast \& Catchpole 1997, Reid 1997) have brought
 down the Cepheid scale distance, thus reducing the estimated Globular
 Cluster age to $t_0 > 12 Gyr$ and $h$ to  a mid term $h=.65$.
  These values still require a cosmological constant, but not as high
   as in $\Lambda$CDM models.
 A value of $\oml > 0.4 $ is needed to satisfy the new age constraint.

 An upper limit on $\Omega_M$ can be obtained from the estimated
 baryonic content of galaxy clusters. If clusters are massive enough
 to represent a fair sample of the total matter content in the universe,
 as numerical simulations confirm (\cite{evr96}), then the baryon fraction
 $f_b=(0.06 \pm 0.003) h^{-3/2}$  (Evrard 1997) should be close to
 its universal value.

 Thus the standard Big Bang Nucleosynthesis value
 $\Omega_b=(0.008-0.024)h^{-2}$ (\cite{cop95}) can be used
  to infer $\Omega_M=\Omega_b /f_b = (0.25 \pm 0.15) h^{-1/2}$.
 For $h=0.65$ one obtains $ \Omega_M < 0.5$.
 For a flat model this limit already overlaps the lower limit from SN Ia.
 This is the main argument for a cosmological constant, possible
  counter arguments like a magnetic field pressure or
  density inhomogeneities, which can lead to underestimate the
  total cluster mass,
 are unlikely to push the limit up to $\Omega_M=1$
 ( Evrard 1997 and references cited therein ).

 The region of parameter space which is allowed by these
observational constraints for a flat model
is then: $ 0.5 \le h \le 0.7$, $ 0.3 \le \Omega_M \le 0.6$.
These are also the limits that for MLM models the cosmological
parameters must independently satisfy
in order to achieve consistency with the set of linear 
clustering data previously analysed.

We think that this is a notable feature
of MLM models  and one
of the most important results of this paper.
We then summarize our conclusions by saying that MLM models
\footnote{While this work was being completed, a paper appeared on babbage
by Eisenstein \& Hu (1997), who have considered 
 power spectra for CDM and other DM models, including MLM.}
appear to be a promising class of cosmological dark matter models.
Our linear analysis shows that  consistency for
the cosmological parameters  
is achieved over a wide range of observational data.

\bigskip
\bigskip
\bigskip

{\bf Acknowledgements}

T. Kahniashvili is grateful to ICTP and SISSA 
for financial support and B. Novosyadlyj
 also acknowledges financial support by  SISSA.
T.K. and B.N. are grateful to SISSA for hospitality and the 
  stimulating academic atmosphere which allowed this work
to progress.
RV thanks also S. Bonometto for helpful discussions.

\bigskip

\begin{thebibliography}{}
\bibitem[Achilli, Occhionero \& Scaramella 1985]{ach85} Achilli, S., 
 Occhionero, F., \&  Scaramella, R. 1985, \apj~299, 577 
\bibitem[Bahcall \& Soneira 1983 ]{bah83} Bahcall, N.A., \& Soneira, R.M. 
1983, \apj~ 270, 20 
\bibitem[Bahcall 1988]{bah88} Bahcall, N. 1988, ARA\&A~ 26, 631
\bibitem[Bahcall 1996]{bah96} Bahcall, N. 1996, astro-ph/9612046
\bibitem[Bahcall \& Cen 1993]{bah93} Bahcall, N.A., \& Cen, R. 1993, \apjl~
 407, L49
\bibitem[Bahcall, Fan \&Cen 1997] {bah97} Bahcall, N., Fan, X., Cen, R. 1997,
 \apjl~ 485, L53
\bibitem[Bardeen, Bond \& Efstathiou 1987 ]{bar87} Bardeen, J.M., Bond J.R., 
 \& Efstathiou, G. 1987, \apj~ 321, 28
\bibitem[Bardeen et al. 1986]{bar86} Bardeen, J.M., Bond,  J.R., Kaiser, N., 
\& Szalay,  A.S. 1986, \apj~ 304, 15 
\bibitem[Bartlett \& Silk 1993]{bar93}   Bartlett, J., G., \& Silk, J. 1993,
\apjl~ 407, L45
\bibitem[Baugh \& Efstathiou 1994]{bau94} Baugh, C.M., \& Efstathiou, G.
 1994, \mnras~ 267, 32
\bibitem[Bennett et al. 1994]{ben94} Bennett, C.L., et al. 1994, \apj ~
    436, 423 
\bibitem[Bennett et al. 1996]{ben96} Bennett, C.L., et al. 1996, \apjl~ 
    464, L1
\bibitem[Bertschinger et al. 1990]{ber90} Bertschinger, E., Dekel, A., Faber, S.M., 
    Dressler A., \& Burstein  D. 1990, \apj~ 364, 370
\bibitem[Bunn \& White 1997]{bun97} Bunn, E.F., \& White, M. 1997,
    \apj~ 480, 6
\bibitem[Bond \& Szalay 1983]{bon83} Bond, J.R., \& Szalay, A.S. 1983, 
    \apj~ 274, 443
\bibitem[Copi, Schramm \& Turner 1995]{cop95} Copi, C.J., Schramm, D.N.,
\& Turner, M.S., 1995, \apj~ 455, L95
\bibitem[Carroll, Press \& Turner 1992]{car92} Carroll, S.M., Press, W.H., \& 
Turner, E.L. 1992, \aap~ 30, 499
\bibitem[Cen \& Ostriker 1994]{cen94} Cen, R., \& Ostriker, J.P. 
    1994, \apj~ 431, 451 
\bibitem[Chaboyer et al. 1996 ]{cha96}  Chaboyer, B. , Demarque, P.,  
Kernan, P.J., \& Krauss, L.M., Science~ 271, 957
\bibitem[Cole et al. 1997]{col97} Cole, S., Weinberg, D.H., Frenk, C.S.,
Ratra, B. 1997, \mnras~ 289, 37
\bibitem[Corteau et al. 93]{cor93} Courteau S., Faber S.M., Dressler A. 
\& Willik J.A. 1993, \apj~ 412, L51
\bibitem[Dalton et al. 1991]{dal91} Dalton, G.B., Efstathiou, G., Lubin, 
  P.M. \& Meinhold, P.R. 1991, PRL~ 66, 2179
\bibitem[Davis et al. 1985]{dav85} Davis, M., Efstathiou, G.,,Frenk, C.S., 
    \& White, S.D.M. 1985, \apj~ 292, 371
\bibitem[Davis \& Efstathiou 1988]{dav88} Davis, M., \& Efstathiou, G. 
    1988,  Large-Scale Motions in the Universe, 
    Rubin, V.C., \& Coyne, G.V., Princeton Univ.Press, Princeton
\bibitem[Davis, Summers \& Schlegel 1992]{dav92} Davis, M., Summers, F.J., 
    \& Schlegel, D. 1992, Nature~ 359, 393 
\bibitem[Dekel et al. 1989]{dck89} Dekel, A., Blumenthal, G.R., Primack, J.P.,
\& Olivier, S. 1989, \apj~ 338, L5
\bibitem[Dekel 1994]{dek94} Dekel, A., 1994, ARA\&A~ 32, 371
\bibitem[Edge et al. 1990]{edg90} Edge, A.C.,  Stewart, G.C., Fabian, A.C., \&
Arnaud, K.A., 1990, \mnras~ 245, 559
\bibitem[Edge \& Stewart 1991]{edg91} Edge, A.C.,  Stewart, G.C., 1991, \mnras ~252, 428
\bibitem[Efstathiou  et al. 1988]{efs88} Efstathiou, G., Frenk, C.S.,
 White, S.D.M., \& Davis, M.,  1988, \mnras ~ 235, 715
\bibitem[Eisenstein \& Hu 1997]{ehu97} Eisenstein, D.J. \& Hu, W., 1997, astro-ph/9710252.
\bibitem[Eke, Cole \& Frenk 1996]{eke96} Eke, V.R. , Cole, S., \& Frenk, C.S.,
 1996, \mnras ~ 282, 263
\bibitem[Eke, Navarro \& Frenk 1997]{eke97} Eke, V.R. , Navarro, J., \& 
Frenk, C.S., 1997, astro-ph/9708070
\bibitem[Evrard, Metzler \& Navarro 1996]{evr96} Evrard, A.E., Metzler, C.A.,
\& Navarro, J.F., 1996, \apj~ 469, 494
\bibitem[Evrard 1997]{evr97} Evrard , A.E., 1997 , astro-ph/9701148 
\bibitem[Fan, Bahcall \&Cen 1997] {fbc97} Fan, X., Bahcall, N., Cen, R., 1997,
astro-ph/9709265
\bibitem[Fang, Xiang \& Li 1984]{fan84} Fang, L.Z., Xiang, S.P., \& Li, S.X., 
    1984, \aap~ 140, 77
\bibitem[Feast \& Catchpole 1997]{fea97} Feast, M.W., \& Catchpole, R.W., 
 1997 , \mnras~ 286 , L1
\bibitem[Frenk et al. 1990]{fre90} Frenk, C., White, S.D.M., \&
    Davis, M. 1990, \apj~ 351, 10
\bibitem[Freedman et al. 1994]{fre94} Freedman, W., et al. 1994, Nature ~
    371, 757 
\bibitem[Haehnelt 1995]{hae95} Haehnelt, M.G., 1995, \mnras 273~, 249
\bibitem[Henry 1997]{hen97} Henry, J.P, 1997, \apj 489~, L1
\bibitem[Henry \& Arnaud 1991]{hen91} Henry, J.P, \& Arnaud K.A. 1991
    \apj 372~, 410
\bibitem[Holtzman 1989]{hol89} Holtzman, J.A., 1989, \apjs~ 71, 1
\bibitem[Holtzman \& Primack 1993]{hol93} Holtzman, J.A., \& Primack , J.R.,
 1993, \apj~ 405, 428
\bibitem[Jing et al. 1993]{jin93} Jing, Y.P., Mo H.J., Borner, G., \& Fang, 
    L.Z. 1993, \apj~ 411, 450
\bibitem[Jing \& Valdarnini 1993]{jiv93} Jing, Y.P., \& Valdarnini, R. 
    1993, \apj~ 406, 6
\bibitem[Klypin \& Rhee 1993]{klr93} Klypin, A.A., \& Rhee, G. ,  
    1993, \apj~ 428, 399
\bibitem[Klypin et al. 1993]{kly93} Klypin, A., Holtzman, J., Primack, 
     J., \& Regos, E. 1993, \apj~  416, 1
\bibitem[Klypin et al. 1995]{kly95} Klypin, A., Borgani, S., Holtzman, J., \& 
Primack, J., 1995, \apj~ 441, 1
\bibitem[Klypin, Primack \& Holtzman 1996]{kly96} Klypin, A., Primack, J., 
    \& Holtzman, J., 1996, \apj~ 466, 13
\bibitem[Klypin \& Kopylov 1983]{kly83} Klypin, A.A., \& Kopylov, A.I, 
    1983, Soviet Astron. Letters 9, 75
\bibitem[Kochanek 1996] {koh96} Kochanek, C.S., 1996,\apj ~ 466, 638
\bibitem[Kofman, Gnedin \& Bahcall 1993]{kof93} Kofman, L.A., Gnedin, 
    N.Y., \& Bahcall, N.A. 1993, \apj~ 413, 1
\bibitem[Kofman \& Starobinsky 1985]{kof85} Kofman, L.A., \& Starobinsky, A.A. 
    1985, SvA Lett.~ 9, 643 
\bibitem[Lacey \& Cole 1994 ]{lac94} Lacey, C., \& Cole, S., 1994, \mnras ~
 271, 676
 \bibitem[Lahav et al. 1989]{lah89} Lahav, O., Edge, A.C., Fabian, A.C. 
 \& Putney, A.  1989, \mnras~ 238, 881
\bibitem[Lahav et al. 1991]{lah91} Lahav, O., Rees, M.J., Lilje, P.B. \& 
    Primack, J. 1991, \mnras~ 251, 128
\bibitem[Lanzetta, Wolfe \& Turnshek 1995]{lan95} Lanzetta, K.M., Wolfe, A.M. 
\& Turnshek, D.A.  1995, \apj~ 440, 435
\bibitem[Liddle \& Lyth 1993]{lid93}Liddle, A.R., \& Lyth, D.H. 1993, 
    Phys. Rep.~ 231, 1
\bibitem[Liddle et al. 1996a]{lia96} Liddle, A.R., Lyth, D.H., Roberts, 
    D. \& Viana, P. 1996a, \mnras~ 278, 644
\bibitem[Liddle et al. 1996b]{lib96} Liddle, A.R., Lyth, D.M., Viana, P.,  
    \& White, M. 1996b, \mnras~ 282, 281
\bibitem[Liddle et al. 1996c]{lic96} Liddle, A.R., Lyth, D.H., Schaefer, 
R.K.,  Shafi, Q., \& Viana, P. 1996c, \mnras~ 281, 531
\bibitem[Liddle et al. 1996d]{lid96} Liddle, A.R., Lyth, D.H., Roberts, 
    D. \& Viana, P. 1996d, \mnras~ 278, 644
\bibitem[Lynden-Bell 1991]{lyn91} Lynde-Bell, D., 1991, in "Observational
 Tests of Cosmological Inflation", ed. by T. Shanks et al., Kluver Academic
 Publishers, vol. 348, p.337.
\bibitem[Ma 1996]{maa96} Ma, C.-P., 1996, \apj~ 471, 13
\bibitem[Ma \& Bertschinger 1994]{maa94} Ma, C.-P., \& Bertschinger, E.,1994, 
    \apjl~ 434, L5
\bibitem[Ma et al. 1997]{mae97} Ma, C.-P., Bertschinger, E., 
Hernquist, L., Weinberg, D.H. \& Katz, N., 1997, \apj~ 484 , L1
\bibitem[McKee et al. 1980]{mck80} McKee, J.D., Mushotzsky, R.F.,
    Boldt, E.A., Holt, S. S., Marchall, F.E.,
    Pravdo, S. H., \& Serlemitsos P.,J. 1980, \apj~ 242, 843
\bibitem[Mo, Jing \& Borner 1993]{moa93} Mo, H.J., Jing, Y.P., \& Borner, 
    G. 1993, \mnras~ 260, 121 
\bibitem[Mo \& Miralda-Escude 1994]{moa94} Mo, H.G. \& Miralda-Escud\'{e}, 1994, 
\apj~ 430, L25
\bibitem[Monaco 1995]{mon95} Monaco, P., 1995, \apj~ 447, 23
\bibitem[Navarro, Frenk \& White 1995]{nav95} Navarro, J.F., Frenk, C.S.,
\& White, S.D.M., 1995, \mnras~ 275, 720 
\bibitem[Novosyadlyj \& Gnatyk 1994]{ng94} Novosyadlyj, B., \& Gnatyk, B.
 1994, Bull. Spec. Astrophys. Obs.~ 37, 81.
\bibitem[Novosyadlyj 1994]{nov94} Novosyadlyj, B. 1994, Kinematics Phys.
 Celest. Bodies~ 10, N1, 7.
\bibitem[Novosyadlyj 1996]{nov96} Novosyadlyj, B. 1996, Astron. \& Astroph.
 Transect.~ 10, 85.
\bibitem[Olivier et al. 1993]{oli93} Olivier S., Primack J., Blumental 
    G.R., and Dekel A., 1993, \apj~ 408, 17 
\bibitem[Peacock 1997]{pea97} Peacock, J.A. 1997, \mnras~ 284, 885
\bibitem[Peacock \& Dodds 1994]{pea94} Peacock, J.A., \& Dodds, S.J. 1994, 
\mnras~ 267, 1020
\bibitem [Perlmutter et al. 1997]{per97} Perlmutter, S., et al. 1997, \apj, 
 to be published, astro-ph/9608192
\bibitem[Peebles 1984]{pee84} Peebles, P.J.E. 1984, \apj~ 284, 439 
\bibitem[Plionis 1995]{pl95} Plionis, M., 1995, in "Clustering in the
 Universe", ed. by S. Maurogordato et al., Editions Frontieres, Singapore,
 p. 273.
\bibitem[Plionis \& Valdarnini 1991]{pv91} Plionis, M., Valdarnini, R., 1991,
  \mnras~ 249, 46.
\bibitem[Pogosyan \& Starobinsky 1993 ]{pog93} Pogosyan, D.Yu. \& Starobinsky, 
    A.A. 1993, \mnras~ 265, 507
\bibitem[Pogosyan \& Starobinsky 1995]{pog95} Pogosyan, D.Yu. \& Starobinsky, 
    A.A. 1995, \apj~ 447, 465
\bibitem[Postman et al. 1992]{pos92} Postman, M., Huchra, J.P. \& Geller,
M.J. 1992, \apj~ 394, 404
\bibitem[Press \& Schechter 1974 ]{pss74} Press  W.H., \& Schechter, P., 1974,
ApJ, 187~, 425
\bibitem[Primack 1997]{pri97} Primack, J.R., 1997, astro-ph/9707285
\bibitem[Primack \& Klypin 1996]{pri96} Primack, J.R., \& Klypin A., 1996, 
astro-ph/9607061
\bibitem[Primack et al. 1995]{pre95} Primack, J.R., Holtzman, J.,
 Klypin A.,  Caldwell, D.O., 1995, PRL~ 74, 2160
\bibitem[Reid 1997]{rei97} Reid, I.N., astro-ph/9704078
\bibitem[Reiss, Kirshner \& Press 1995]{rie95} Reiss, A.G., Kirshner, R.P., \&
Press, W.H., 1995, \apjl~  438, L17
\bibitem[Shafi \& Stecker 1984]{sha84} Shafi, Q., \& Stecker, F.W. 1984, 
    P.R.L.~ 53, 1292
\bibitem[Schaefer \& Shafi 1992]{sch92} Schaefer, R.K., \& Shafi, Q. 1992, 
    Nature~ 359, 199
\bibitem[Schuster et al. 1993]{SP93} Schuster, J., Gaier, T., Gundersen, J.
 et al., 1993, \apj~ 412, L47.
\bibitem[Smoot et al. 1992]{smo92} Smoot, G.F., Bennett, C.L., Kogut, A. et 
    al. 1992, \apjl~ 396, L1
\bibitem[Squires at al. 1996]{squ96} Squires at al. 1996, \apj~ 461, 572
\bibitem[Stompor, Gorsky \& Banday 1995]{sto95} Stompor, R., Gorski K.M., \& 
    Banday L.  1995, \mnras~ 277, 1225
\bibitem[Storrie-Lombardi et al. 1996]{sto96} Storrie-Lombardi, L.J., 
Mc Mahon, R.G., Irwin, M.J. \& Hazard, C., 1996, \apj~ 468, 121
\bibitem[Sugiyama  1995]{sug95} Sugiyama, N., 1995, \apjs~ 100, 281
\bibitem[Taylor \& Rowan-Robinson 1992]{tay92} Taylor, A.N. \& Rowan-Robinson, 
    M. 1992, Nature~ 359, 396
\bibitem[Valdarnini \& Bonometto 1985]{val85} Valdarnini, R., \&
    Bonometto, S.A. 1985, \aap~ 146, 235
\bibitem[Van Dalen \& Schaefer 1992]{van92}Van Dalen , A., \& Schaefer, R.K. 
1992 , \apj~ 398, 33
\bibitem[Viana \& Liddle 1996]{via96} Viana, P.T.P., \& Liddle, A.R. 1996,
    \mnras~ 281, 323 
\bibitem[Walker et al. 1991]{wal91} Walker, T.P., Steigman, G., Schramm, 
    D.N., Olive, K.A., \& Kang, H.S. 1991, \apj~ 376, 51
\bibitem[White, Frenk \& Davis 1983]{whi83} White, S.D.M., Frenk, C.S., \&
 Davis, M. 1983 , \apjl~ 274, L1
\bibitem[White, Efstathiou \& Frenk 1993a]{wha93} White, S.D.M., Efstathiou, 
    G., \& Frenk, C.S. 1993a , \mnras~ 262, 1023
\bibitem[White et al. 1993b]{whb93} White, S.D.M., Navarro, J.F, Evrard, A.E.,
\& Frenk, C.S. 1993b , Nature~ 366, 429
\bibitem[Wolfe 1993]{wof93} Wolfe, A., 1993, in {\it Relativistic Astrophysics
and Particle Cosmology }, eds.  C.W., Ackerlof, M.A., Srednicki 
( New York: New York Academy of Science ) , p.281
\bibitem[Zakharov 1979]{zak80} Zakharov A.V. 1979, Zh. Eksp. Teor. Fiz.
(USSR) (Sov. Phys. JETF)~ 77, 434
\bibitem[Zamorani et al. 1991]{zam91} Zamorani, G., Scaramella, R.,
Vettolani, G. \& Chincarini, G. 1991, Proceedings of the Workshop on
'Traces of the primordial Structure in the Universe', May 1991, ed.
Bohringer, H. \& Treumann, R.A., MPE report 227, p.59
\end{thebibliography}
\end{document}